\documentclass[aps,preprintnumbers,pra,showpacs,twocolumn]{revtex4-1}

\preprint{LMU-ASC 14/15}

\usepackage{slashed}
\usepackage{mathrsfs}
\usepackage{amsfonts}
\usepackage{amsmath}
\usepackage{amssymb}
\usepackage{revsymb}
\usepackage{graphicx}
\usepackage{mathrsfs}
\usepackage{bm}
\usepackage{bbm}
\usepackage{psfrag}
\usepackage{color}
\usepackage[]{units}

\usepackage[hidelinks]{hyperref}

\usepackage{natbib}
\usepackage[usenames,dvipsnames]{xcolor}
\usepackage{dashrule}
\newcommand{\ampp}{\mathcal{E}_{p}}
\newcommand{\amps}{\mathcal{E}_{s}}

\newcommand{\bea}{\begin{eqnarray}} 
\newcommand{\eea}{\end{eqnarray}} 
\newcommand{\eps}{\varepsilon}
\newcommand{\veps}{\pmb{\varepsilon}}
\newcommand{\vepspar}{\pmb{\varepsilon}^{\parallel}}
\newcommand{\vepsperp}{\pmb{\varepsilon}^{\perp}}
\newcommand{\mbf}[1]{\mathbf{#1}}
\newcommand{\trm}[1]{\textrm{#1}}
\newcommand{\tsf}[1]{\textsf{#1}}

\newcommand{\figref}[1]{Fig. \ref{#1}}

\newcommand{\eqnref}[1]{Eq. (\ref{#1})}
\newcommand{\eqnrefs}[2]{Eqs. (\ref{#1}) and (\ref{#2})}
\newcommand{\eqnreft}[2]{Eqs. (\ref{#1}-\ref{#2})}
\newcommand{\sxnref}[1]{Sec. \ref{#1}}
\newcommand{\appref}[1]{App. \ref{#1}}
\newcommand{\tabref}[1]{Tab.$~$\ref{#1}}

\newcommand{\Ecr}{E_{\trm{cr}}}

\newcommand{\vphi}{\varphi}

\newcommand{\rect}{\trm{Rect}}

\newcommand{\es}{\pmb{\eps}_{s}}
\newcommand{\ep}{\pmb{\eps}_{p}}

\newcommand{\nref}{\tsf{n}}

\newlength\replength
\newcommand\repfrac{.33}

\newcommand\tdotfill[1][\repfrac]{\cleaders\hbox to \replength{%
  \smash{\raisebox{\arraystretch\dimexpr\ht\strutbox-.1ex\relax}{.}}}\hfill}

\begin{document}
\title{Vacuum high harmonic generation in the shock regime}

\author{P. \surname{B\"ohl}}
\email{patrick.boehl@physik.uni-muenchen.de}
\affiliation{Arnold Sommerfeld Center for Theoretical Physics, \\ 
Ludwig-Maximilians-Universit\"at M\"unchen,
    Theresienstra\ss e 37, 80333 M\"unchen, Germany}

\author{B. \surname{King}}
\email{b.king@plymouth.ac.uk}
\affiliation{
    Centre for Mathematical Sciences, Plymouth University, 
    Plymouth PL4 8AA, United Kingdom}

\author{H. \surname{Ruhl}}
 \email{hartmut.ruhl@physik.uni-muenchen.de}
\affiliation{Arnold Sommerfeld Center for Theoretical Physics, \\ 
Ludwig-Maximilians-Universit\"at M\"unchen,
    Theresienstra\ss e 37, 80333 M\"unchen, Germany}

\date{\today}
\begin{abstract}
Electrodynamics becomes nonlinear and permits the self-interaction of fields when the quantised 
nature of vacuum states is taken into account. The effect on a plane probe pulse 
propagating through a stronger constant crossed background is calculated using numerical 
simulation and by analytically solving the corresponding wave equation. The electromagnetic 
shock resulting from vacuum high harmonic generation is investigated and a nonlinear 
shock parameter identified. 
\end{abstract}
\maketitle

%
%
%
%
\section{Introduction}
Soon after the formulation of relativistic quantum mechanics, it became clear that the propagation 
of light through the vacuum would be modified due to the polarisability of virtual 
electron-positron pairs \cite{sauter31, halpern34, euler35, weisskopf36}. Heisenberg and Euler 
derived the Lagrangian of an effective description of this interaction for constant 
fields \cite{heisenberg_euler36}, which was later rederived by Schwinger \cite{schwinger51}. 
Derivative expansions of this effective interaction \cite{gusynin96, gusynin99, dunne99} and 
numerical worldline calculations \cite{gies11} imply that ``constant'' is to be
taken with respect to the Compton time $h/mc^{2}$ for electron mass $m$. This suggests a good 
approximation of the effect for time-dependent fields with a much longer period than the Compton 
time is to simply insert them in place of the constant fields in the Heisenberg-Euler Lagrangian. 
In 
particular, the 
polarised vacuum supports the phenomenon of 
self-interaction when two electromagnetic waves couple via virtual electron-positron pairs and 
the 
principle of superposition no longer holds.
\newline

There have been several studies of the consequences of this self-interaction. Lutzky and Toll 
\cite{toll59}
showed that if the field invariant $\mathcal{G}=-FF^{\ast}/4\Ecr^{2}=\mbf{E}\cdot\mbf{B}=0$ 
where $F$, $F^{\ast}$ are the Faraday tensor and its dual, $\Ecr = 
m^{2}c^{3}/\hbar e=1.3\cdot10^{16}\,\trm{Vcm}^{-1}$ is the so-called ``critical'' 
field, $e>0$ is the charge of a positron and $\mbf{E}$ and $\mbf{B}$ the total electric 
and magnetic fields in units of the critical field, a current that depends nonlinearly on the 
invariant 
$\mathcal{F}=-F^{2}/4\Ecr^{2}=(E^{2}-B^{2})/2$ leads to the generation of an electromagnetic 
discontinuity 
or ``shock''. After identifying an application in magnetised neutron stars, shocks
were analysed in a constant magnetic field background 
using a first- \cite{rozanov93}, second- \cite{fabrikant82} and
several-  \cite{heyl98,heyl99} order weak-field expansion of the Heisenberg-Euler Lagrangian with 
an 
all-order analysis performed by Bialynicka-Birula \cite{bb81}. An astrophysical environment was 
further modelled by introducing nonlinear vacuum effects into equations of relativistic 
magnetohydrodynamics \cite{heyl99} and into a dusty plasma \cite{marklund05}. 
\newline

In the current article we analyse a pump-probe set-up of having a linearly-polarised oscillating 
plane wave (probe) counterpropagate through a linearly-polarised constant crossed and stronger 
plane wave 
background (pump). Of particular interest will be the two cases of having parallel or 
perpendicular probe and pump 
polarisations. Observables 
are 
expressed in terms of the electric and magnetic fields to aid comparison with numerical simulation. 
\newline

Unlike in classical electrodynamics where a superposition of solutions to the wave equation is also 
a solution, when the existence of charged virtual electron-positron vacuum states is included, the 
principle of superposition is no longer valid. A consequence of using the Heisenberg-Euler 
Lagrangian is that a non-trivial vacuum ``current'' appears in Maxwell's equations, which 
disappears 
in the classical limit $\hbar \to 0$. If the electromagnetic fields are not very weak $E \gg 
\sqrt{\alpha}(\hbar \omega/mc^{2})^{2}$ \cite{narozhny15} where $\alpha\approx\nicefrac[]{1}{137}$ 
is the fine-structure constant and the field frequency is
$\omega$ (corresponding to 
an 
intensity much greater than $10^{5}\,\trm{Wcm}^{-2}$ for an optical 
laser), they can be regarded as classical. When this is the case, the methods of classical 
electrodynamics 
can be used to solve Maxwell's equations with the vacuum current. 
\newline

\begin{figure}[!h] 
\centering
\includegraphics[draft=false,width=8cm]{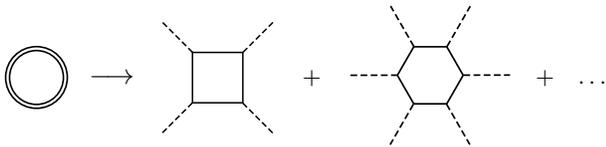}
 \caption{An illustration of the weak-field expansion of the vacuum polarisation diagram.}
 \label{fig:HE_WF} 
\end{figure}

For fields much weaker than critical, the interaction with the virtual electron-positron pairs of 
the vacuum permits $2n$-wave mixing for integer $n>1$ such as four- and six-wave mixing, as 
demonstrated in \figref{fig:HE_WF}.
One can make an analogy with nonlinear optics, in which the polarisation $\mbf{P}$ of an optical 
material can depend upon higher powers of the electric field \cite{franken63}, which are described 
using different 
orders of the susceptibility tensor $\chi^{(j)}$:
\bea
P_{i} = \chi^{(1)}_{ij} E_{j} + \chi^{(2)}_{ijk}E_{j}E_{k} + \ldots,
\eea
and analogously for the magnetisation $\mbf{M}$. Being a relativistic effect however, the magnetic 
and the 
electric field appear in the vacuum polarisation and magnetisation on an equal footing. For weak 
fields and propagation lengths shorter than the scattering length, four-wave mixing 
is the most probable vacuum polarisation process for colliding plane waves (with the exception of 
certain special field geometries). This is often compared to the optical Kerr effect 
\cite{marklund10}, but in a pump-probe experiment in which the probe oscillates much quicker 
than the pump field, the steepening of the carrier wave \cite{rosen65} and not 
the envelope \cite{demartini67} is stronger.
\newline

If the fields' spacetime extent is much larger than a single scattering length, multiple 
$2n$-wave mixing can occur, in which the change in field due to wave mixing influences further 
changes due to wave 
mixing, with each mixing event involving a potentially different $n$. Usually it is 
assumed that the probability for multiple mixing events is much lower than single mixing 
events, and multiple events are neglected. However, if the extent of 
the field is large enough, 
this hierarchy can be broken and it can become more probable that multiple mixing 
events occur 
than a single mixing event so that all orders of wave mixing events have to 
be taken into account. With the generation of a large number of higher harmonics, the shape 
of the electromagnetic plane waves will also change and this leads to the possibility of shock wave 
generation. In the ``shock regime'', as all orders of wave mixing can play a role in the 
generation of the spectrum, the spectrum is expected to be qualitatively different to the 
perturbative case of having only a single mixing event, where four-wave mixing is the most 
probable and higher harmonics are exponentially suppressed. Such a type of shock generation is 
also known from nonlinear optics \cite{anderson82}.
\newline

In contrast to this, the weak-field expansion in \figref{fig:HE_WF} suggests that high harmonic 
generation can also occur through single scattering events that involve large numbers of photons. 
The likelihood of this happening increases with the field strength of slowly-varying weak fields. 
This type of \emph{vacuum 
high harmonic generation} has been investigated using the full 
polarisation operator in \cite{dipiazza05, fedotov_harmonics06} and using the lowest order of the 
weak-field expansion in \cite{lundstroem_PRL_06,king12,boehl14,karbstein14a}. A highlight of 
the current article is the first investigation of vacuum high harmonic 
generation in plane wave fields in what 
we call the \emph{shock regime}, where the probe propagation length is much larger than the mean 
scattering length. It will be shown that in certain parameter regimes, this can be a much more 
efficient high harmonic generation mechanism.
\newline

The aims of this work are: i) to investigate vacuum high harmonic generation in the collision of 
plane waves that are weaker than critical, for the case that the fields' spacetime extent is 
much larger than the mean scattering length; ii) to show that the higher harmonics are accompanied 
by an electromagnetic shock due to the polarised vacuum; iii) to investigate the 
dependency of this shock on the colliding fields' mutual linear polarisation; iv) to comment on the 
similarities and differences of high harmonic generation in laser-irradiated plasmas.
\newline

We begin with a derivation of the modified Maxwell and wave equations (Sec. 
II), summarise the 
analytical method (Sec. III) and the numerical method used in computational simulation (Sec. 
IV) before analysing higher harmonic generation with just four-wave mixing (Sec. 
V), just six-wave mixing (Sec. VI) and both four- and six-wave mixing (Sec. VII). 
We then discuss 
the results, compare with high harmonic generation from oscillating plasmas (Sec. VIII) and 
conclude (Sec. IX).

%
%
%
%
%
\section{Modified Electromagnetic Wave Propagation}
The Heisenberg-Euler Lagrangian can be written \footnote{This corrects the extra factor of $\alpha$ 
in Eq. (4) from \cite{boehl14}} \cite{schwinger51}
\begin{align}
\mathcal{L}_{\trm{HE}} &=& 
-\frac{m^{4}}{8\pi^{2}}\int_{0}^{\infty}\!\!ds\,\frac{\mbox{e}^{-s}}{s^{3}}
\Big[s^{2}ab\,\trm{cot}\,as\,\trm{coth}\,bs - 1 \nonumber\\
&&\qquad\qquad\qquad\qquad+\frac{s^{2}}{3}(a^{2}-b^{2})\Big],\label{eqn:LEHfull}
  \end{align}
(we have set here and throughout $\hbar=c=1$ unless they explicitly occur), where the secular 
invariants $a$ and $b$ are given 
by:
\begin{align}
  a =
  \left[\sqrt{\mathcal{F}^{2}+\mathcal{G}^{2}}+\mathcal{F}\right]^{1/2};
  &\quad& b =
  \left[\sqrt{\mathcal{F}^{2}+\mathcal{G}^{2}}-\mathcal{F}\right]^{1/2} \nonumber
\end{align}
and we recall that electric and magnetic fields are in units of the critical field $\Ecr$. 
Applying the Euler-Lagrange equations to 
$\mathcal{L}=\mathcal{L}_{\trm{MW}}+\mathcal{L}_{\trm{HE}}$, where
$\mathcal{L}_{\trm{MW}}=m^4(E^2 - B^2)/8\pi\alpha$ is the classical Maxwell Lagrangian, gives 
the modified Maxwell equations:
\begin{align}
 \partial_{\mu}F^{\ast\mu\nu}=0\label{eqn:MWEa}\\
 (1+C_{1})\,\partial_{\mu}F^{\mu\nu}
 +C_{2}\,F^{\mu\nu}\partial_{\mu}F^{2}
 +C_{3}\,F^{\ast\mu\nu}\partial_{\mu}(FF^{\ast})
 \nonumber\\ 
 +~C_{4}\,\Big[ F^{\ast\mu\nu}\partial_{\mu}F^{2}
 +F^{\mu\nu}\partial_{\mu}(FF^{\ast})\Big]
 = 0 \label{eqn:MWE}
 \end{align}
and the general expressions for the coefficients $C_i$ are given in \appref{sxn:Cs}.
Expressing these equations in electric and magnetic fields, we acquire:
\bea
\nabla\wedge\mbf{E}+\partial_{t}\mbf{B}&=&0 \label{eqn:MW1}\\
\nabla\wedge\mbf{B}-\partial_{t}\mbf{E}&=&\mbf{J}[\mbf{E},\mbf{B}] \label{eqn:MW2}
\eea
\begin{align}
\begin{split}
\mbf{J}[\mbf{E},\mbf{B}] = \big[& C_1 \left(\partial_t \mbf{E} - \nabla \wedge \mbf{B}\right)  
+\left(C_{2}\mbf{E}+C_{4}\mbf{B}\right)\partial_{t}F^2\\
&+\left(C_{2}\mbf{B}-C_{4}\mbf{E}\right)\wedge\nabla F^2\\
&+\left(C_{4}\mbf{B} -C_{3}\mbf{E}
\right)\wedge\nabla (FF^{\ast})\\
& +\left(C_{3}\mbf{B}+C_{4}\mbf{E}\right)\partial_{t}(FF^{\ast})\big]. \label{eqn:current_maxwell}
\end{split}
\end{align}
The current $\mbf{J}$ in Maxwell's equations is related to the 
corresponding source $\mbf{T}$ in the wave equation for the electric field via $\partial_{t} 
\mbf{J} = \mbf{T}$. 
\newline

We restrict our analysis to the case when $E \ll 1$, for two reasons. First, it 
allows us to 
neglect 
the creation of real electron-positron pairs, as the probability of vacuum pair production in a 
volume equal to the reduced Compton wavelength $\lambdabar = \hbar/mc$ cubed in the 
Compton time 
$\lambdabar/c$ is $P = 
E^{2}\exp(-\pi/ E)/4\pi^{3}$ \cite{schwinger51}, which is heavily suppressed for $E\ll 1$. Second, 
it permits a 
perturbative expansion in $E$, the so-called ``weak-field expansion'', of the Heisenberg-Euler 
Lagrangian. 
\newline

Although all electromagnetic fields are classical, it is useful to envisage the 
corresponding quantum process involving photons and this is depicted for the weak-field expansion 
of the vacuum polarisation operator in \figref{fig:HE_WF}. (Indeed, it has been shown that the 
leading-order term of the weak-field expansion agrees with the direct calculation of the 
four-photon box diagram in the low-frequency limit $\hbar\omega\ll mc^{2}$ 
\cite{karplus50}.) The weak-field expansion of \eqnref{eqn:LEHfull} for $E\ll1$ is then:
 \begin{align}
 \mathcal{L}_{\trm{HE}} &= \frac{m^{4}}{\alpha}\sum_{i=1}^{\infty}  
\mathcal{L}_{i},\label{eqn:LEHwf}\\
 \mathcal{L}_{1} &=  \frac{\mu_{1}}{4\pi}\left[\left(E^{2}-B^{2}\right)^{2} +  
7(\mbf{E}\cdot\mbf{B})^{2}\right]\label{eqn:L1},\\
 \mathcal{L}_{2} &=  \frac{\mu_{2}}{4\pi}\left(E^{2}-B^{2}\right)\left[2\left(E^{2}-B^{2} 
\right)^{2} +13\left(\mbf{E}\cdot\mbf{B}\right)^{2} \right], \label{eqn:L2}\\
\mathcal{L}_{3} &=  \frac{\mu_{3}}{4\pi}\left[3\left(E^{2}-B^{2}\right)^{4} 
+22\left(E^{2}-B^{2}\right)^{2}\left(\mbf{E}\cdot\mbf{B}\right)^{2}\right.  \nonumber\\
& \qquad +\left. 19\left(\mbf{E}\cdot\mbf{B}\right)^{4} \right],\label{eqn:L3}
 \end{align}
where $\mu_{1}=\nicefrac[]{\alpha}{90\pi}$, $\mu_{2}=\nicefrac[]{\alpha}{315\pi}$, 
$\mu_{3}=\nicefrac[]{4\alpha}{945\pi}$ 
(although $\alpha$ 
occurs in the denominator in \eqnref{eqn:LEHwf}, as fields are in units of the critical field, when 
$\hbar\to 0$, $\mathcal{L}_{\trm{HE}}\to0$). The coefficients $C_i$
in \eqnref{eqn:MWE} that follow from  $\mathcal{L}_1$ and $\mathcal{L}_2$ 
are given by
\eqnreft{eqn:C1b}{eqn:C4h} in \appref{sxn:Cs}.
\newline

In the scenario we consider, the initial electric field is $\mbf{E}^{(0)}(\vphi_{p},\vphi_{s}) = 
\mbf{E}^{(0)}_{p}(\vphi_{p}) + \mbf{E}^{(0)}_{s}(\vphi_{s})$ and the initial probe and strong 
electric waves are given by:
\bea
\mbf{E}^{(0)}_{p}(\vphi_{p}) & = & 
\veps_{p}\,\mathcal{E}_{p}\,\mbox{e}^{-\left(\frac{\vphi_{p}}{\Phi_{p}}\right)^{2}}\cos\vphi_{p} 
\label{eqn:Ep0}\\
\mbf{E}^{(0)}_{s}(\vphi_{s}) & = & 
\veps_{s}\,\mathcal{E}_{s}\,\rect\left(\frac{\vphi_{s}}{\Phi_{s}}\right), \label{eqn:Es0}
\eea
where the rectangular function $\rect(\vphi/\Phi) = \theta(\vphi + \Phi/2) - \theta(\vphi-\Phi/2)$ 
and $\theta(\cdot)$ the Heaviside function \cite{wang12},  $\vphi_{p} = k_{p}x = 
\omega_{p}x^{-}$, $\vphi_{s} = k_{s}x= \omega_{s}x^{+}$, $x^{\pm}=t\pm z$, 
$\Phi_{p}=\omega_{p}\tau_{p}$, $\Phi_{s}=\omega_{s}\tau_{s}$ with the 
probe and strong field polarisation vectors $\veps_{p}$, $\veps_{s}$ 
obeying $\veps_{p}\cdot\veps_{p}=1$, 
$\veps_{s}\cdot\veps_{s}=1$, $\mbf{k}_{p}\cdot\veps_{p} = 0$, $\mbf{k}_{s}\cdot\veps_{s} = 0$ and 
the probe pulse is assumed to be much weaker than the strong background $\mathcal{E}_{p} \ll 
\mathcal{E}_{s}$. Initially, the probe and strong fields are well separated: 
$\lim_{t\to-\infty}\mathcal{F}, \mathcal{G} = 0$. We define the 
orthonormal polarisation vectors $(\vepspar,\vepsperp)$ where $\vepspar \equiv \veps_{p}$ 
defines ``parallel'' polarisation, and $\vepsperp$ ``perpendicular'' polarisation with
$\vepsperp\cdot\vepspar=0$, $\vepsperp\cdot\mbf{k}_{p}=0$.
\newline

Since the vacuum current is a function of the relativistic invariants $\mathcal{F} = 
(E^{2}-B^{2})/2$ and 
$\mathcal{G} = \mbf{E}\cdot\mbf{B}$, for single plane waves, there is no effect on propagation 
due to vacuum polarisation \cite{schwinger51,king12b,king12c}. Therefore, the only contributions 
will come from 
cross-terms between the probe and strong field. As the weak-field expansion is an expansion in 
powers of $\mathcal{F}$ and $\mathcal{G}$, for our scenario, each order scales as 
$\mathcal{L}_{n}\sim (E_{s}E_{p})^{n+1}$.
\newline

The initial probe 
$\mbf{E}_{p}^{(0)}$ and strong $\mbf{E}_{s}^{(0)}$ fields satisfy the classical vacuum wave 
equation independently:
\bea
\square\, \mbf{E}_{p}^{(0)} = \pmb{0}~, \qquad\square\, \mbf{E}_{s}^{(0)} = \pmb{0}, \nonumber
\eea
where $\square = c^{-2}\partial_{t}^{2}-\nabla^{2}$.
The effect of the polarised vacuum can be included with a source term 
$\mbf{T}=\mbf{T}[\mbf{E},\mbf{B}]$ occurring on the right-hand side of the wave equation. 
We will assume that solutions to this equation are also plane waves propagating along the 
same axis as the pump and probe waves. This allows us to write $\mbf{T}=\mbf{T}[\mbf{E}]$. 
Since a single plane wave cannot polarise the vacuum \cite{king12b, king12c}:
\bea
\mbf{T}[\mbf{E}_{p}] = \pmb{0}~, \qquad \mbf{T}[\mbf{E}_{s}] = \pmb{0}. \nonumber
\eea 
However, since two counterpropagating plane waves \emph{can} polarise the vacuum, the wave equation 
we will solve is:
\bea
\square\, \left(\mbf{E}_{p} + \mbf{E}_{s}\right) = \mbf{T}[\mbf{E}_{p}+\mbf{E}_{s}]. 
\label{eqn:pwe1}
\eea
In particular, we are interested in solutions which include the self-action of the probe that 
lead to a plasma-like vacuum instability and corresponding electromagnetic shock. \eqnref{eqn:pwe1} 
will be solved in two ways. First, the scattered probe will be solved for using an analytical 
method based on an 
iterative procedure that ignores changes to the stronger background:
\bea
\square\, \mbf{E}_{p}^{(n+1)} = \mbf{T}[\mbf{E}_{p}^{(n)}+\mbf{E}_{s}^{(0)}]. \label{eqn:itwe}
\eea
Second, \eqnref{eqn:pwe1} will be solved consistently in a numerical simulation that uses tools 
based on the pseudocharacteristic method of lines, which are applied to the corresponding Maxwell 
equations. In this way, the ``asymptotic'' state of the probe field after it has passed 
through the strong field and $\mbf{T}\approx \pmb{0}$ (in contrast to the ``overlap'' dynamics when 
$\mbf{T} \neq \pmb{0}$ \cite{boehl14}) will be studied.
\newline

As we are considering the collision of counter-propagating plane waves, the general Maxwell's 
equations in \eqnrefs{eqn:MW1}{eqn:MW2} reduce to one spatial ($z$) and one temporal ($t$) 
dimension. 
To determine which terms in the full weak-field expansion for the current 
\eqnref{eqn:LEHwf} should be considered when calculating high harmonic generation, we employ the 
following scaling argument. As explained in \cite{boehl14}, the change in the field due to 
interaction with the vacuum that propagates with the probe (``forward'' scattering) is 
\bea
\Delta \mbf{E}_{p}(x^{-}) = \int_{-\infty}^{z}\!\frac{dz'}{2}~\mbf{J}(t'=x^{-}+z',z'), 
\label{eqn:J_use}
\eea
where the vacuum current is:
\bea
\mbf{J} =\sum_{i=1}^{\infty}\mbf{J}_{i}~;\qquad \mbf{J}_{i} = 
4\pi\left[\mbf{\widehat{k}}_{p}\wedge 
\partial_{z}\mbf{M}_{i} + \partial_{t}\mbf{P}_{i}\right]. \label{eqn:jdef}
\eea
$\mbf{\widehat{k}}_{p}=\mbf{k}_{p}/|\mbf{k}_{p}|$ and the dimensionless vacuum polarisation 
$\mbf{P}_{i} = 
\partial \mathcal{L}_{i}/ \partial \mbf{E}$ and magnetisation 
$\mbf{M}_{i} = \partial \mathcal{L}_{i}/ \partial \mbf{B}$ (as used in e.g. \cite{bb81}  or 
\cite{dipiazza06}). The forward-scattered signal is zero if the vectorial part of $\mbf{P}_{i}$ or 
$\mbf{M}_{i}$ is from the probe field. As already explained, 
$\mathcal{L}_{n}\sim(E_{s}E_{p})^{n+1}$, but in the wave equation that results from this, the 
vacuum current contains derivatives with respect to $E_{s}$ and $E_{p}$. Since the current 
containing the derivative with respect to $E_{s}$ vanishes for forward scattering in plane waves 
\cite{adler71, boehl14}, we see that the remaining current and hence the scattered field 
$J_{n} \propto 
\mu_{n}\mathcal{E}_{s}^{n+1}\mathcal{E}_{p}^{n}$. The integration over $z'$ is over the strong 
field 
and so contributes a factor $\tau_{s}$ and the differentials in \eqnref{eqn:jdef} contribute 
approximately a 
factor $\omega_{p}$, so that one can estimate $\Delta E_{p}^{(1)} \propto 
\mu_{n}\mathcal{E}_{s}^{n+1}\mathcal{E}_{p}^{n}\Phi$, for $\Phi = \omega_{p}\tau_{s}$. Since we 
assume $E \ll 1$, and since we are interested in the case when the change in the probe is 
of the same order as the probe field and self-interaction becomes important, we require $\Phi \gg 
1$. We 
also note that the coefficients $\mu_{n}$ diverge with $n$ because the weak-field expansion is 
asymptotic (see e.g. \cite{dunne04b}), so we do not expect the series can be truncated for 
arbitrarily 
large $n$ and still yield a useful approximation. Although purely four-photon scattering does 
allow the generation of higher harmonics in this set-up, this first occurs for double four-photon 
scattering. The 
contribution from this twice-iterated process appears in $\Delta \mbf{E}_{p}^{(2)}$ and 
scales as $\propto(\mu_{1})^{2}\mathcal{E}_{s}^{3}\mathcal{E}_{p}^{2}\Phi$, which when compared to 
the leading 
contribution to second harmonic generation from six-photon scattering in $\Delta 
E_{p}^{(1)}\propto\mu_{2}\mathcal{E}_{s}^{3}\mathcal{E}_{p}^{2}\Phi$, is suppressed by a factor 
$(\mu_{1})^{2}/\mu_{2} \ll 1$. Therefore, when considering higher harmonic generation along the 
probe propagation axis in the  
regime 
$E \ll 1$, $\Phi \gg 1$, the leading contribution originates from six-photon scattering. In Sec. V, 
this simple scaling argument will be seen to agree with the full numerical analysis. An argument 
for 
neglecting eight-photon scattering will be forthcoming.
%
%
%
%

\section{Analytical Method}\label{sec:am}
To solve the inhomogeneous wave equation
\bea
\left[\partial_{t}^{2}-\partial_{z}^{2}\right] \mbf{E}_{p} = 
\mbf{T}[\mbf{E}_{p}+\mbf{E}_{s}^{(0)}], 
\label{eqn:we1}
\eea
we employ an iterative ansatz:
\bea
\mbf{E}_{p}^{(n+1)} = \mbf{E}_{p}^{(0)} + \Delta\mbf{E}_{p}^{(n)}, \label{eqn:itan}
\eea
where
\bea
\Delta\mbf{E}_{p}^{(n)}(t,z) = \int dt'\,dz' G(t-t',z-z')\mbf{T}^{(n)}(t',z'), \nonumber
\eea
and in general
\bea
\mbf{T}^{(n)}(t,z) = 
\sum_{i=1}^{\infty}\mbf{T}_{i}\left[\mbf{E}_{p}^{(n-1)}(\vphi_{p})+\mbf{E}_{s}^{(0)}(\vphi_{s}
)\right ] , \nonumber
\eea
where the subscript $i$ is the order of the weak-field expansion and the retarded Green's function 
is \cite{mahan02}:
\bea
G(t,z) = \frac{\nref}{2}\theta(t)\theta\left(\frac{t}{\nref}-|z|\right), \nonumber
\eea
for refractive index $\nref$. If $\nref=1$, one acquires \eqnref{eqn:J_use}, where 
$\partial_{t}\mbf{J}^{(n)}(t,z) = \mbf{T}^{(n)}(t,z)$. These equations can be iterated to 
calculate the generation of higher harmonics due to multiple 
scattering as outlined in the introduction. Within this analytical approach, we assume 
$\omega_{p}\tau_{p} \gg 1$ and $\omega_{p}\tau_{s} \gg 1$, so that the derivative of the probe 
and background envelopes can be neglected with respect to the derivative of the oscillating part of 
the probe in $\mbf{J}$. When studying the generation of higher harmonics, we will be particularly
interested in taking
\bea
\mbf{T}^{(n)}(t,z) = \mbf{T}_{2}\left[\mbf{E}_{p}^{(n-1)}(\vphi_{p})+\mbf{E}_{s}^{(0)}(\vphi_{s}
)\right], \nonumber
\eea
which corresponds to considering purely six-photon scattering (this will be further justified 
shortly).

\begin{figure}[!h] 
\centering
\includegraphics[draft=false,width=6.75cm]{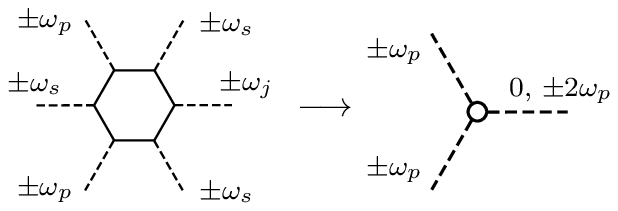}
 \caption{In the left-hand diagram, $\omega_{j} \in 
\{\omega_{s},3\omega_{s},2\omega_{p}\pm\omega_{s},2\omega_{p}\pm3\omega_{s}\}$. If the 
strong field is approximated as constant and the three strong-field photon legs are
suppressed, in an effective approach, six-photon scattering of the probe can be represented as a 
triple interaction. The $\pm$ refer to incoming and outgoing photons respectively.}
 \label{fig:hex_eff_diag} 
\end{figure}

A diagrammatic approach is useful to understand the physical processes described by different 
iterations of the probe field $\mbf{E}_{p}^{(n)}$. First, since we are interested in harmonic 
generation and since the background is constant, we suppress strong-field photon legs. Furthermore, 
as the Heisenberg-Euler Lagrangian is ``effective'' in that all fermion dynamics have been 
integrated out, all vacuum loops are reduced to effective vertices. Then the diagram representing 
six-photon scattering, which is the leading order harmonic-generating process, is given in 
\figref{fig:hex_eff_diag}.
The iterative ansatz in \eqnref{eqn:itan} is 
illustrated in \tabref{tab:its}. 
\begin{center}
\begin{table}
{\setlength{\tabcolsep}{8.0pt}
\begin{tabular}{rc}
\multicolumn{2}{c}{$\phantom{.}$ \hspace{-0.6cm}\rule{8.4cm}{0.02cm}}\\[-2.25ex]
\multicolumn{2}{c}{$\phantom{.}$ \hspace{-0.6cm}\rule{8.4cm}{0.02cm}}\\[1ex]
\multicolumn{2}{c}{\raisebox{-.5\height}{\includegraphics[draft=false]{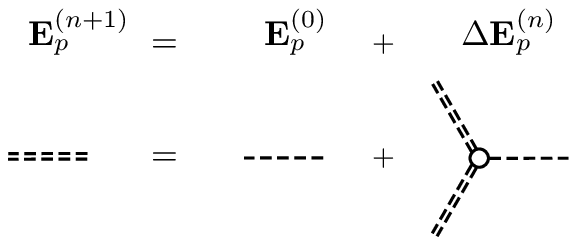}}}\\[3ex]
\multicolumn{2}{c}{$\phantom{.}$ \hspace{-0.6cm}\rule{8.4cm}{0.02cm}}\\[1ex]
$\mbf{E}_{p}^{(0)}~$: & 
\raisebox{-.5\height}{\includegraphics[draft=false]{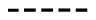}}\\[3ex]
$\mbf{E}_{p}^{(1)}~$: & 
\raisebox{-.5\height}{\includegraphics[draft=false]{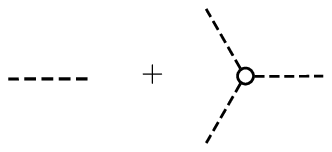}}\\[3ex]
$\mbf{E}_{p}^{(2)}~$: & 
\raisebox{-0.79\height}{\includegraphics[draft=false]{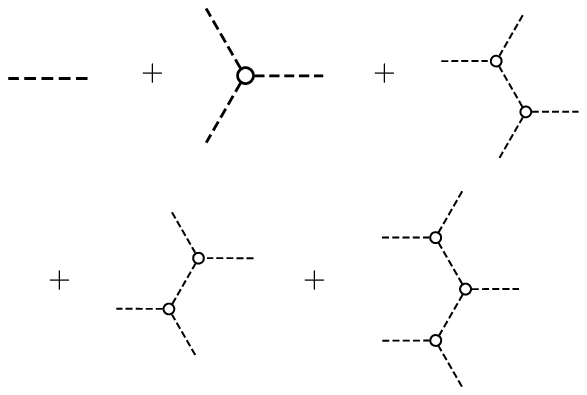}}\\
\multicolumn{2}{c}{$\phantom{.}$ \hspace{-0.6cm}\rule{8.4cm}{0.02cm}}\\[-2.25ex]
\multicolumn{2}{c}{$\phantom{.}$ \hspace{-0.6cm}\rule{8.4cm}{0.02cm}}\\
\end{tabular}
}
\caption{Diagrammatic representation of the first iterations of the probe wave equation.} 
\label{tab:its}
\end{table}
\end{center}

The diagrammatic equation in \tabref{tab:its} in some ways
resembles the Schwinger-Dyson equation \cite{gribov01} but in this case the left-hand side is the 
self-consistent solution of the probe field at a particular 
order of iteration, and the double line on the right-hand side is where the scattered probe field 
from the previous order is applied.
In \tabref{tab:its}, it is shown how the 
number of diagrams rapidly increases with iteration order (as the square of the number in the 
previous 
order plus one, although many are equivalent). It also demonstrates that terms of a much higher 
perturbative order (number of vertices) are generated at a given iterative order ($\mbf{E}^{(n)}$ 
contains terms from the $(2^{n}-1)$th perturbative order, but is only accurate to the $n$th 
perturbation order). 
\newline

\begin{figure}[!h] 
\centering
\includegraphics[draft=false, width=4cm]{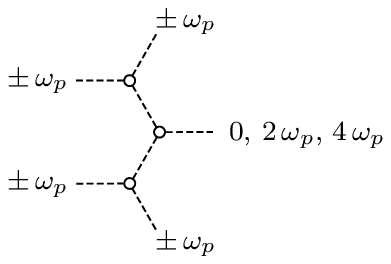}
\caption{An example of the harmonics generated in the probe due to effective self-interaction in a 
slowly-varying background.}
\label{fig:it_ex} 
\end{figure}

On all the diagrams with at least one vertex, one leg is the scattered field and the rest 
are 
incoming or outgoing probe photons. An example is given in \figref{fig:it_ex} where the $\pm$ sign 
refers to the energy added to the system by incoming/outgoing photons. 
By summing the series that occurs in $\lim_{n\to\infty}\mbf{E}_{p}^{(n)}$, we will arrive at an 
analytical 
expression for the asymptotic probe field and in doing so identify a shock parameter that signifies 
when self-action effects become important.
\newline

\begin{figure}[!h] 
\centering
\includegraphics[draft=false,width=6cm]{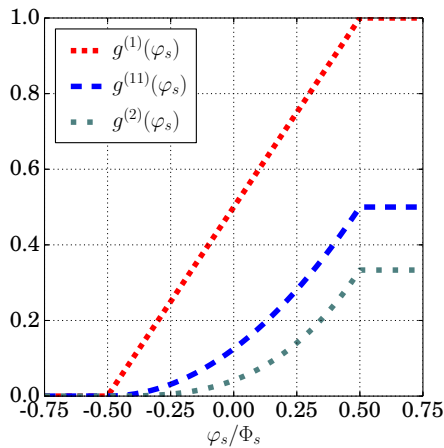}
 \caption{(Color online) A plot of how the functions describing how the occurrence of higher harmonics varies with 
probe propagation length.}
 \label{fig:gs} 
\end{figure}

For the example of parallel probe and strong field polarisation, the second 
iteration 
shown in \tabref{tab:its} is:
\bea
\mbf{E}_{p}^{(2)} &=& 
\veps_{p}\,\mathcal{E}_{p}\mbox{e}^{-\left(\frac{\vphi_{p}}{\Phi_{p}}\right)^{2}}\left[
\left(1-\left(\frac{v}{2}\right)^ { 2 }
g^{(11)}(\vphi_{s})\right)\cos\vphi_{p}\right.\nonumber\\
&&\left. -\frac{v}{2}g^{(1)}(\vphi_{s})\sin2\vphi_{p} - 
3\left(\frac{v}{2}\right)^{2}g^{(11)}(\vphi_{s})\cos3\vphi_{p}\right.\nonumber \\
&& \left. +2\left(\frac{v}{2}\right)^{3}g^{(2)}(\vphi_{s})\sin4\vphi_{p}\right], 
\label{eqn:secit}
\eea
where $v=\nu_{2}\exp(-(\vphi_{p}/\Phi_{p})^{2})$ and the shock parameter 
$\nu_{2} = 192\mu_{2}\mathcal{E}_{s}^{3}\mathcal{E}_{p}\Phi$. The functions of $\vphi_{s}$ 
describe how the particular 
term is generated \emph{during} the passage of the probe through the strong background (all fields 
are classical) and 
originate 
from repeated integration of the interaction over co-ordinate. Here:  
\bea
g^{(1)}(\vphi_{s}) &=& \int_{-\infty}^{\vphi_{s}/\Phi_{s}}\!dy~\rect(y)\label{eqn:g1_function}\\
g^{(11)}(\vphi_{s}) &=& \int_{-\infty}^{\vphi_{s}/\Phi_{s}}\!dy~\rect(y)~g^{(1)}(y)\nonumber\\
g^{(2)}(\vphi_{s}) &=& 
\int_{-\infty}^{\vphi_{s}/\Phi_{s}}\!dy~\rect(y)\,\left[g^{(1)}(y)\right]^{2}, \nonumber
\eea
and these are plotted in \figref{fig:gs}.

As mentioned in the introduction, we are mainly interested in the \emph{asymptotic} state of 
the probe:
\bea
\mbf{E}_{p}^{(2)}(\vphi_{p}) = \lim_{\vphi_{s}\to\infty}\mbf{E}_{p}^{(2)}(\vphi_{p},\vphi_{s}),
\eea
where we note $\lim_{\vphi_{s}\to\infty}g^{(1)}(\vphi_{s}) = 1$, 
$\lim_{\vphi_{s}\to\infty}g^{(11)}(\vphi_{s}) = \nicefrac[]{1}{2}$ and 
$\lim_{\vphi_{s}\to\infty}g^{(2)}(\vphi_{s}) = \nicefrac[]{1}{3}$.
\newline

As previously remarked, using this method, $\mbf{E}_{p}^{(n)}$ contains powers of $v$ from $0$ to 
$2^{n}-1$ but is only 
accurate to $O(v^{n})$. We also note that the $n$th iteration generates harmonics from $1$ to 
$2^{n}$. A power series in $v$ multiplies each harmonic so we can write a given iteration as:
\bea
\mbf{E}_{p}^{(n)}(\vphi_{p},\vphi_{s}) &=& 
\veps_{p}\,\mathcal{E}_{p}\mbox{e}^{-\left(\frac{\vphi_{p}}{\Phi_{p}}\right)^{2}}\sum_{j=1}^{\infty}
\left[a^{(n)}_{2j}(v,\vphi_{s})\sin2j\vphi_{p}
\right.\nonumber \\
&& \left. \qquad + a^{(n)}_{2j-1}(v,\vphi_{s})\cos(2j-1)\vphi_{p}\right]. 
\label{eqn:Epn}
\eea
Of most interest is the asymptotic state of the full solution:
\bea
\mbf{E}_{p}(\vphi_{p}) = \lim_{\vphi_{s}\to\infty} \lim_{n\to \infty} 
\mbf{E}_{p}^{(n)}(\vphi_{p},\vphi_{s}), \nonumber
\eea
and we find that for the parallel set-up:
\bea
\lim_{\vphi_{s}\to\infty}\lim_{n\to \infty}a_{j}^{(n)}(v,\vphi_{s}) =a_{j}(v)= 2(-1)^{\lfloor 
\nicefrac[]{j}{2}
\rfloor}\frac{J_{j}(jv)}{jv},\nonumber \\ \label{eqn:harmonicCoeffs}
\eea
where $\lfloor j \rfloor = \trm{floor}(j)$ and $J_{l}(\cdot)$ is the $l$th-order Bessel function 
of the first kind \cite{gradshteyn07}. We note that the all-order solution 
\eqnref{eqn:harmonicCoeffs} for a plane probe propagating through a constant crossed, 
parallel-polarised 
background, resembles the Fubini solution \cite{fubini35} for the propagation of lossless 
finite-amplitude planar acoustic waves in nonlinear media \cite{rossing07}.
\newline

The all-order solution can be derived from a probe-dependent refractive index: 
$\nref=1+\delta \nref_{2}$ with $\nu_{2}=\delta\nref_{2}\Phi$ where:
\bea
\delta\nref_{2}(\vphi_{s},\vphi_{p}) = 192\mu_{2}E_{s}^{3}(\vphi_{s})E_{p}(\vphi_{p}).
\eea
So the scattered probe field due to just six-photon scattering can be written:
\bea
\mbf{E}_{p}(\vphi_{p}) = 
\mbf{E}_{p}^{(0)}\left(\vphi_{p}+\nu_{2}[E_{p}(\vphi_{p})]\right).
\eea
\newline

To justify when it is a good approximation to only consider six-photon scattering, let us consider 
first eight-photon scattering. The shock-parameter for eight-photon scattering is
$\nu_{3}=1536\mu_{3}\mathcal{E}_{s}^{4}\mathcal{E}_{p}^{2}\Phi$. In order that this is much 
less 
than $\nu_{2}$, we require $\mathcal{E}_{s}\mathcal{E}_{p} \ll \nicefrac[]{3}{32}$, and 
since 
$\mathcal{E}_{s}\ll1$ and $\mathcal{E}_{p}\ll1$, this is fulfilled. Therefore the individual 
effect of the next higher-order terms in the weak-field expansion should be negligible. In 
contrast, the importance of four-photon scattering can be quantified by the parameter $\upsilon_{1} 
= 16\mu_{1}\mathcal{E}_{s}^{2}\Phi$ but this corresponds to the process of one incoming and one 
outgoing photon from a scattering event and therefore will not contribute directly to harmonic 
generation. Nevertheless, it does lead to a refractive index alteration, which in combination with 
multiple six-photon scattering, could potentially influence the generated spectrum. To 
ignore this in our analysis would require $\mathcal{E}_{s}\mathcal{E}_{p} \gg 
\nicefrac[]{7}{24}$, which 
is 
not fulfilled. To explore this point, the simulation 
results are split into three cases: i) purely four-photon scattering, ii) purely six-photon 
scattering
and iii) both four- and six- photon scattering. We consider two polarisation scenarios: the 
``parallel set-up'' and the ``perpendicular set-up'', 
which refer to the initial strong field polarisation being in the $\vepspar$ and $\vepsperp$ mode 
respectively. In the parallel set-up, we will find all harmonics are generated in the parallel 
polarisation mode $\vepspar$, whereas for the perpendicular set-up, each odd harmonic will be 
generated in a perpendicular mode $\vepsperp$ and each even harmonic in a parallel one $\vepspar$.
%
%
%
%
\section{Numerical Method}
For the scenario of two colliding plane wave pulses, the modified Maxwell equations in
\eqnrefs{eqn:MW1}{eqn:MW2} can be written in matrix form: 
  \bea  
\left(\mathbbm{1}_{4}+\mbf{X}\right)\partial_{t}\mbf{f}+\left(\mathbf{Q}+\mbf{Y}\right)\partial_{z}
\mbf{f}=0,\label{eqn:MatrixForm1D}
  \eea
  where $\mbf{f} = (E_{x},E_{y},B_{x},B_{y})^{T}$, $\mathbbm{1}_{4}$ is the identity matrix in four 
dimensions, 
$\mbf{Q}=\trm{adiag}(1,-1,-1,1)$ is an anti-diagonal matrix and $\mbf{X}$ and $\mbf{Y}$ 
are the perturbations due to vacuum 
interaction given in a general form in \appref{sec:matricesAB}.
\newline  

Our numerical method, which was first employed by the authors in
\cite{boehl14} and will be explained in more detail in the following,
is based on inverting the matrix $(\mathbbm{1}_{4}  + \mbf{X})$ to convert 
\eqnref{eqn:MatrixForm1D} to
 a system of ordinary differential equations (ODEs), discretising in
 space using the ``pseudocharacteristic 
method of  lines'' (PCMOL) \cite{carver1980pseudo} and integrating the equations of motion using 
the ODE solver CVODE  \cite{Hindmarsh:2005:SSN:1089014.1089020}.
\newline

Our analysis is valid when $\mathcal{E}_{s}, \mathcal{E}_{p} \ll 1$ and the single parameter 
relevant to high harmonic generation in the shock regime that 
depends on the field strength is $\nu_{2}=192\mu_{2}\mathcal{E}_{s}^{3}\mathcal{E}_{p}\Phi$. We 
wish to simulate the occurrence of a shock wave, for which $\nu_{2}\to 1$, implying $\Phi$ 
must be very large in order to compensate for the weak field strengths. However, a large $\Phi$ is 
computational expensive to simulate. To compare analytical and numerical results, we will therefore 
extrapolate the theoretical result to values of $\mathcal{E}_{s} \not \ll 1$, allowing a simulation 
for smaller $\Phi$ to be performed, with the condition that the physical prediction is only valid 
for a particular value of $\nu_{2}$ when $\mathcal{E}_{s} \ll 1$. For this reason, we will often 
quote 
simulation parameters in terms of shock parameters rather than absolute field strengths and 
spatial extensions. 

\subsection{Linear case}
Let us first consider (\ref{eqn:MatrixForm1D}) with 
$\mbf{X}=\mbf{Y}=\pmb{0}$, which is the $\hbar\to0$ limit.
This system is hyperbolic \cite{strikwerda2004finite}, which means that we can find a basis
$\mbf{u}:=\mbf{S}\,\mbf{f}$ such that the matrix $\mbf{\Lambda}= 
\mbf{S}\mbf{Q}\mbf{S}^{-1}=\mathrm{diag}(-1,-1,1,1)$  is diagonal with real eigenvalues:
\begin{align}
  \mbf{S}\!=\!\frac{1}{\sqrt{2}}
  \left( \begin{smallmatrix}
      \text{-}1&   0& 0& 1\\
      0&   1& 1& 0\\
      1&   0& 0& 1\\
      0& \text{-}1& 1& 0
    \end{smallmatrix} \right)
  \quad
  \mbf{u}\! :=\! 
  \mbf{S}\,\mbf{f}=\!\frac{1}{\sqrt{2}}\left(\begin{smallmatrix}
      B_y-E_x\\ E_y+B_x\\ E_x+B_y\\ B_x-E_y 
    \end{smallmatrix}\right).\label{eq:Ubasis}
\end{align}
In this new basis, we have an uncoupled system of advection
equations:
\begin{equation*}
  \partial_{t}\mbf{u}(t,z)+\mbf{\Lambda}\,\partial_{z}\mbf{u}(t,z)=0.
  \label{eqn:MatrixFormLinear1D}
\end{equation*}
The diagonal elements $\lambda_i$ of $\mbf{\Lambda}$ are called the ``characteristic speeds`` 
the system, where  $\lambda_{i}=\pm1$ corresponds to a component travelling along the 
characteristics 
$x^{\pm}$ 
with the speed of light.  We proceed by introducing a 
co-located grid for the components $u_i$ with $N$ grid points.  The field components $u_i$ on 
the grid are arranged blockwise in a 
large $4N$-dimensional vector $\mbf{\tilde u}=(\ldots u^{l-1}_4u^l_1u_2^lu_3^lu_4^l u_1^{l+1}\ldots)$, 
where $u_i^l=u_i(l\Delta 
z)$ and $0<l\leq N$ is the index of the grid point. The PCMOL uses biased
differencing for each component $u_i$ according to the sign of the corresponding characteristic 
speed
$\lambda_i$, where the component $u_i$ with $\lambda_i>0$ ($\lambda_i
<0$) is thereby differentiated using backward (forward) finite differences using fourth-order 
accuracy. In \cite{schiesser1991numerical} it is argued that
  this biased differencing using five-point-stencils is an
  effective fixed grid method for first order hyperbolic partial differential equations
  because it shows a good balance between introducing minimal numerical
  diffusion and oscillations in the solution where steep gradients are
  present.
The derivatives at the boundary are also approximated using only field values inside the box. 
Instead of transforming the system 
back to $\mbf{\tilde f}$ (the tilde in this section indicates the discretised version 
on the grid), which is normally done in the PCMOL, the system is solved for $\mbf{\tilde u}$. This 
has the advantage of having open boundary conditions 
since the components $u_i$ are only allowed to flow in one direction. 
If we take the system to be of size $L$ and a spatial resolution of $N$ grid points,
 then distance is measured in units of $\Delta z = \nicefrac[]{L}{(N-1)}$, where $N-1$ corresponds
 to the boundary conditions being taken into account.
We are left with a system of ODEs $\mbf{\tilde u}'(t) = 
\mbf{g}[\mbf{\tilde u}(t),t]$, 
where $\mbf{g}[\mbf{\tilde u}(t),t]=-\mbf{\tilde \Lambda}\,\mbf{D}\,\mbf{\tilde u}$, with the
$4\times 4$ matrix $\mbf{\Lambda}$ being mapped onto a $4N \times 4N$ dimensional block-diagonal one, $\mathbf{\tilde\Lambda} = \mathbbm{1}_{N} \otimes \mathbf{\Lambda}$ ($\otimes$ is the Kronecker product \cite{lancaster1985theory}) and $\mbf{D}$ being the $4N\times 4N$ matrix representing the biased 
differencing explained
above. For the detailed action of $\mbf{D}$ on $\mbf{\tilde u}$ see
  Appendix \ref{sec:stencils}.
\newline

The initial conditions are set up in $\mbf{\tilde f}$, the system is integrated in 
$\mbf{\tilde u}$ using CVODE  and transformed back for output.  CVODE is an 
ODE-solver that offers variable-order, variable-step multi-step methods. 
Initially, we supply the ``right-hand-side function'' $\mbf{g}[\mbf{\tilde u}(t),t]$ as above. 
Since both the linear and nonlinear cases are non-stiff (no rapidly-damped modes are expected), we 
apply the Adams-Moulton-Methods together with the 
variational method to solve the 
resulting linear system. This provides higher accuracy with less computational effort 
compared to 
the offered Newton iterations, since neither approximations nor an analytical expression for the 
Jacobian 
have to be provided. We always use the parallel implementation of CVODE together with  
``extended'' (long double) precision. 

\subsection{Nonlinear case}
By discretising the full nonlinear system (\ref{eqn:MatrixForm1D}), the matrices $\mathbf{X}$ and $\mathbf{Y}$ also become $4N\times 4N$ dimensional. The system then can also 
be brought into ODE form $\mbf{\tilde u}'(t)=\mbf{g}[\mbf{\tilde u}(t),t]$ by inverting the 
matrix $(\mathbbm{1}_{4N}+\mbf{\tilde X})$. Since $\mbf{X}$ depends only on the field components, 
the full matrix can be written as $\mathbf{\tilde X} = \mathbbm{1}_{N} \otimes
  \mathbf{X}^l$  and the upper index denotes the former $4\times 4 $
  matrix $\mathbf{X}$ at grid point $l$. This can be used to reduce
  the inversion of $\mathbf{\tilde X}$ to $N$ times the inversion of a $4\times 4$ matrix. The structure of $\mbf{X}^l$ allows us to rewrite
$\mbf{X}^l$ as $\mbf{X}^l=\mbf{G}\,\mbf{H}^l$ 
with
\begin{equation}
  \mbf{G}=
 \begin{pmatrix}
    1 & 0 \\
    0 & 1 \\
    0 & 0\\
    0& 0
  \end{pmatrix}, \quad \quad\mbf{H}^l = 
  \begin{pmatrix}
     x^l_{11} & x^l_{12} & x^l_{13} & x^l_{14}\\
   x^l_{21} & x^l_{22} & x^l_{23} & x^l_{24}\\
\end{pmatrix}\nonumber
\end{equation}
where the $x^l_{ij}$ are the values of the non-vanishing matrix elements of
  $\mbf{X}$ given in App. \ref{sec:matricesAB}, evaluated at position $l$. Then we can apply the 
Woodbury Formula \cite{golub2012matrix},
\begin{align}
 (\mathbbm{1}_4+\mathbf{X}^l)^{-1} = \mathbbm{1}_4 - \mbf{G}(\mathbbm{1}_2+ 
\mbf{H}^l\mbf{G})^{-1}\mbf{H}^l\ ,\nonumber
\end{align}
to further reduce the inversion to one of the $2\times2$ matrix
\begin{align}
  (\mathbbm{1}_2 +\mbf{H}^l\mbf{G}) = \begin{pmatrix}
    1+x^l_{11} & x^l_{12} \\ 
    x^l_{21}  & 1+x^l_{22}
    \end{pmatrix}\ . \nonumber
  \end{align}
This is performed for all grid points using an LU-factorisation at each
evaluation of the function $\mbf{g}[\mbf{\tilde u}(t),t]$.
\newline

For the parameters considered, the nonlinear corrections $\mbf{X}$ 
and $\mbf{Y}$ do not change the signs of the characteristic speeds, so
we use the same biased differencing as in the linear case.
The nonlinear ODE-system is then given by
  \begin{align}
  \mbf{\tilde u}'(t) =-\mbf{\tilde S}(\mathbbm{1}_{4N} + 
\mbf{\tilde X})^{-1}(\mbf{\tilde Q}+\mbf{\tilde Y})\mbf{\tilde S}^{-1} \mbf{D}\mbf{\tilde u}\ . \nonumber
\end{align}
where $\mathbf{\tilde S} = \mathbbm{1}_{N} \otimes \mathbf{S}$,
$\mathbf{\tilde Q} = \mathbbm{1}_{N} \otimes \mathbf{Q}$ and
$\mathbf{\tilde Y} = \mathbbm{1}_{N} \otimes \mathbf{Y}^{l}$ in analogy to
$\mathbf{\tilde X}$. All fields are normalised by $\Ecr$.
The parameters for CVODE are the same as in the linear case. The signals are analysed under the 
assumption $\omega = |\mbf{k}|$ using a spatial Fourier Transform 
in \textit{Wolfram Mathematica} \cite{mathematica9}.

\subsection{Simulational Setup}
 Recalling the form of the probe and strong pulses (\eqnrefs{eqn:Ep0}{eqn:Es0}), we consider a 
Gaussian probe pulse with base frequency $\omega_p$ and a ``constant'' strong pulse.  We consider 
the two cases of parallel and perpendicular set-ups which are 
characterised by parallel and perpendicular polarisations respectively of the probe and the strong 
pulse with $\pmb{\eps}_p\cdot\pmb{\eps}_s=1$ and $\pmb{\eps}_p\cdot \pmb{\eps}_s=0$. 
The rectangular shape of the strong pulse is approximated using a mirrored Fermi-Dirac distribution 
in the simulation box. The function $\mathrm{FD}(y)$ is given by 
\begin{align} 
\mathrm{FD}(y)=\frac{1}{1+\exp(\frac{|y|- \omega_s z_m}{ \omega_s z_b})} \ . \label{eqn:FD}
\end{align}
The parameters $z_b$ and $z_m$ play the role of the ``temperature'' and ``chemical potential'', 
controlling the steepness and width of the strong pulse. Typical values are 
$z_b= 5 \cdot 10^{-5}\mathrm{cm}$ and $z_m=100\cdot z_b$. 
\newline

\begin{figure}[!h] 
 \includegraphics[draft=false,width=0.79\linewidth]{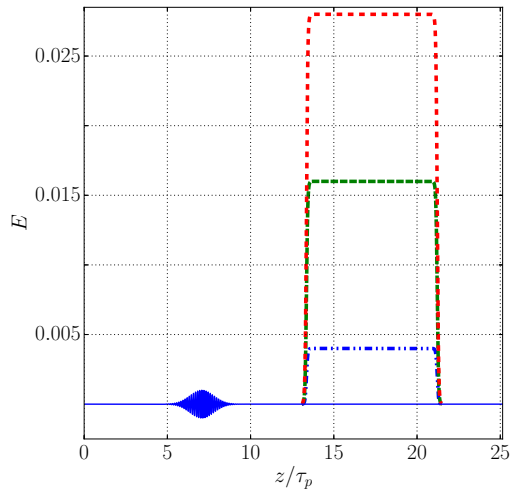}
 \caption{(Color online) The simulational set-up for the collision of a Gaussian probe pulse with a 
``constant'' pulse of various strengths (indicated by different line styles) and identical polarisation. The size of the system is taken to be $3.2\cdot 
10^{-2}\,\trm{cm}$.}
 \label{fig:ccfsetup} 
 \end{figure}

A snapshot of the simulation box for $t=0$ is shown in \figref{fig:ccfsetup}. The use of a 
Fermi-Dirac distribution
  instead of a \rect-function follows the advice in \cite{schiesser1991numerical}, where it is 
recommended to avoid sharp gradients (which is infinite for a
  \rect-function) because of numerical diffusion and spurious
  oscillations. To ensure the accuracy of the simulations, we use a sufficiently high number of grid points for the
  Fermi-Dirac-function and generated shock waves in order to resolve the gradients
  properly so that spurious effects are suppressed.
  
The initial conditions are:
 \begin{align}
   \mbf{E}_p(\phi_p,\phi_{0p}) &= \ep \mathcal{E}_{p}\, 
e^{-\frac{(\phi_p-\phi_{0p})^2}{\Phi_p}}\cos(\phi_p-\phi_{0p} )\nonumber\\[0.8pt]
   \mbf{E}_s(\phi_s,\phi_{0s})  &= \pmb{\eps}_{s} \mathcal{E}_{s} 
\rect((\phi_s-\phi_{0s})/\Phi_s)\nonumber \\[0.8pt]
&\approx \pmb{\eps}_{s} \mathcal{E}_{s} \mathrm{FD}(\phi_s-\phi_{0s})\nonumber\\[0.8pt]  
\mbf{B}_{i}(\phi_i,\phi_{0i})&=\widehat{\mbf{k}}_{i}\wedge\mbf{E}_{i}(\phi_i,\phi_{0i}
)\nonumber
   \end{align}
with $\phi_i =\omega_i z$, $\phi_{0i}=\omega_{i} z_{0i}$, $\Phi_i = \omega_i \tau_i$ and 
$i\in\{p,s\}$.
\newline

 To define the pulse duration $\tau_s$ of 
the strong pulse when using the Fermi-Dirac function, we equate the calculation of the first 
iteration for the simulational parallel setup with 
the analytical model (see \eqnrefs{eqn:J_use}{eqn:g1_function}):
 \begin{align}
    \Delta\mbf{E}^{(1)}_p(\varphi_p) &=\lim_{\varphi_s\rightarrow \infty} -\es\ampp 
e^{-\left(\frac{\varphi_p}{\Phi_p}\right)^{2}}\frac{v}{2}\frac{h^{(1)}(\vphi_{s})}{\tau_{s}}
\sin 
2\varphi_p\nonumber \\
 &=-\es\ampp 
e^{-\left(\frac{\varphi_p}{\Phi_p}\right)^{2}}\frac{v}{2}\sin 
2\varphi_p\nonumber \ ,
 \end{align}
where $h^{(1)}(\varphi_s)$ is given by
\begin{align}
h^{(1)}(\vphi_{s}) = 
\frac{1}{\omega_s}\int_{-\infty}^{\vphi_{s}}\!dy~\mathrm{FD}^3(y)\nonumber \ .
\end{align}
The duration $\tau_s$ is then defined by $\tau_s = \lim_{\varphi_s\rightarrow \infty}
h^{(1)}(\varphi_s)$. The initial conditions are chosen such that the field invariants
 and the field values at the boundary are essentially zero initially and
  the system is simulated until the pulses are again well separated.
  \newline

\begin{figure}[!h]
\includegraphics[width=0.79\linewidth]{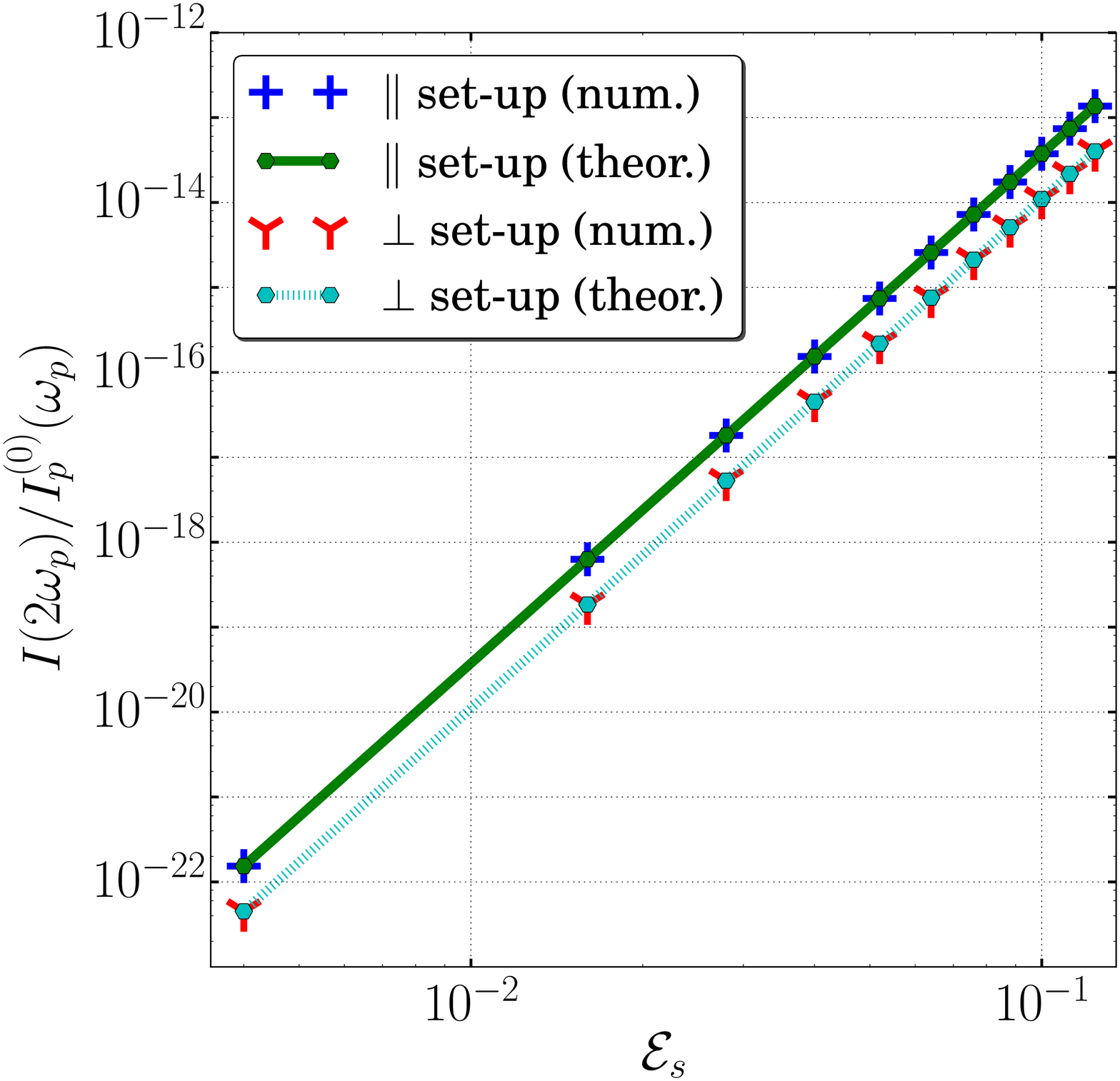}
\caption{(Color online) The relative intensity of the second harmonic generated by single six-photon 
scattering for $\ampp=10^{-3}$.}\label{fig:secondharmonic}
\end{figure}

Results of the simulation were compared to the analytical result 
for asymptotic lowest order second harmonic generation in the parallel
and perpendicular set-ups \cite{boehl14}, where the Gaussian strong
background in \cite{boehl14} is replaced with the mirrored Fermi-Dirac distribution
\eqnref{eqn:FD}. The excellent agreement is displayed in \figref{fig:secondharmonic}, where the 
log-log plot of the 
ratio $I(2\omega_{p})/I_{p}^{(0)}(\omega_{p})$ for various values of the strong field 
amplitude is calculated using:
\begin{gather}
\frac{I(\omega)}{I_{p}^{(0)}(\omega_{p})} = 
\left[\frac{|\widetilde{E}_{p}(\omega)|}{|\widetilde{E}^{(0)}_{p}(\omega_{p})|
} \right ] ^ { 2 }\!,~ \mbf{\widetilde{E}}_{p}(\omega) = 
\int_{-\infty}^{\infty}\!dx^{-}~\mbf{E}_{p}(x^{-})\,\mbox{e}^{i\omega x^{-}}.
\end{gather}

%

%
%
%
%
\section{All-order four-photon scattering} \label{sxn:aofps}
For the parameter regime of interest, the most probable effect on the probe pulse due to 
four-photon scattering is that from the well-studied 
modified vacuum indices of refraction 
$\nref^{\parallel,\perp}_{1}=1+\delta \nref^{\parallel,\perp}_{1}$ 
given by \cite{baier67b,gies00}:
\bea
\delta\nref^{\parallel,\perp}_{1} = 2(11\mp3)\mu_{1}\amps^{2},
\eea
which can be written in a phase-dependent way $\nref_{1}(\vphi_{s})=1+\delta \nref_{1}(\vphi_{s})$:
\bea
\delta \nref_{1}(\varphi_{s}) = 
4\mu_{1}\left[E_{s}^{(0)}(\varphi_{s})\right]^{2}\left[4\left(\pmb{\eps}_{p}
\cdot\pmb { \eps } _ { s } \right)^ { 2 } +7\left(\pmb{
\eps}_{p}\wedge\pmb{\eps}_{s}\right)^{2}\right].\nonumber \\ \label{eqn:dnconst}
\eea
Following the analytical method in \sxnref{sec:am}, summing all perturbative orders, one finds due 
to purely four-photon scattering (corresponding to using $\mbf{T}=\mbf{T}_{1}$ in 
\eqnref{eqn:we1}), in 
the parallel set-up:
\bea
\mbf{E}_{p}(\vphi_{p}) &=& 
\sum_{j=0}^{\infty}\frac{\upsilon_{1}^{j}}{j!}\frac{d^{j}}{d\vphi_{p}^{j}}\mbf{E}_{p}^{(0)}(\vphi_{p
} )= 
\mbox{e}^{\upsilon_{1}\frac{d}{d\vphi_{p}}}\mbf{E}_{p}^{(0)}(\vphi_{p}), \nonumber
\eea
which is just a shift-operator in the phase that is applied to the initial probe pulse giving:
\bea
\mbf{E}_{p}(\vphi_{p}) &=& \mbf{E}_{p}^{(0)}(\vphi_{p}+\upsilon_{1}), \nonumber
\eea
where the multi-scale parameter for the parallel and perpendicular cases $\upsilon_{1} = 
\upsilon_{1}^{\parallel, \perp}$:
\bea
\upsilon_{1}^{\parallel,\perp}=2(11\mp 3)\mu_{1}\mathcal{E}_{s}^{2}\Phi = \delta 
\tsf{n}_{1}^{\parallel,\perp}\Phi .
\eea 
This all-order solution to the phase shift in a 
plane wave propagating through a constant background derived from the Heisenberg-Euler 
Lagrangian complements a recent example solution of the phase shift derived from the 
Schwinger-Dyson equation applied to the polarisation operator \cite{meuren15}.
\newline

\begin{figure}[!h]
\includegraphics[width=0.7\linewidth]{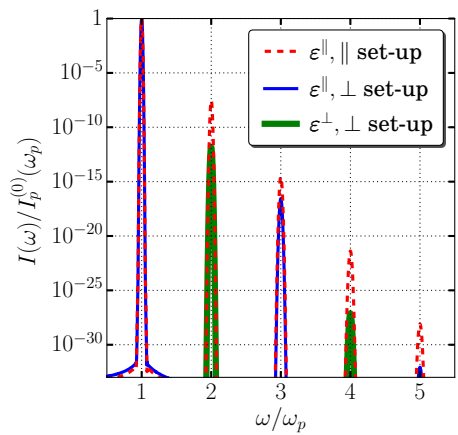}
\caption{(Color online) High harmonic generation from multiple four-photon scattering for 
$\nu_{1}=3.3\times 10^{-4}$.
} \label{fig:box_chained}
\end{figure}

Photon merging via single four-photon scattering is prohibited in a plane wave 
counterpropagating parallel to the background 
\cite{adler71,boehl14}. However, when the possibility of multiple four-photon scattering is taken 
into account, high harmonic 
generation \emph{can} take place. The modified refractive index \eqnref{eqn:dnconst}, experienced 
by the probe due to the strong field  and conversely the modified refractive index 
experienced by the strong field due to the probe, leads to the electromagnetic invariants 
$\mathcal{F}$, $\mathcal{G}$ no longer vanishing for the probe and strong fields separately. A 
log-log plot of the normalised spectrum $I(\omega)/I_{p}^{(0)}(\omega_{p})$ for various cases of 
high harmonic generation through purely four-photon scattering is displayed 
in \figref{fig:box_chained}.
\newline

The perpendicular set-up leads to even harmonics being generated in the $\vepspar$ mode and odd 
harmonics being 
generated in the $\vepsperp$ mode. All higher harmonics in the perpendicular set-up are 
suppressed compared to the parallel set-up, with odd harmonics being suppressed more than even 
ones. In the parallel set-up, all photons are scattered into the $\vepspar$ mode. As 
will become clear in \sxnref{sxn:aosps}, compared to the six-photon channel, harmonic generation 
via four-photon scattering is considerably 
suppressed. The scaling argument given at the end of \sxnref{sec:am} can now be understood in the 
following way. For purely four-photon scattering, one scattering event must have occurred 
to change the electromagnetic variants (a factor $\delta \nref_{1} = 16\mu_{1} 
\mathcal{E}_{s}^{2}$) 
and one further scattering with a probe photon (a factor 
$\upsilon_{1}=16\mu_{1}\mathcal{E}_{s}\mathcal{E}_{p}\Phi$), which yields the combination
\bea
\nu_{1}=(16\mu_{1})^{2}\mathcal{E}_{s}^{3}\mathcal{E}_{p}\Phi. 
\eea
If one takes this to be the 
shock parameter for purely four-photon scattering, for the parameters of the parallel set-up in 
\figref{fig:box_chained}, following \eqnref{eqn:secit}, one would expect the second harmonic at 
relative intensity 
$(\nu_{1}/2)^{2}=10^{-7.6}$, third harmonic at 
$(3\nu_{1}^{2}/8)^{2}=10^{-14.8}$ and the fourth harmonic at 
$(\nu_{1}^{3}/12)^{2}=10^{-23.0}$, which correctly predict the 
numerical results to within an order of magnitude. For comparison, the shock parameter for 
purely six-photon scattering for this set-up would be $\nu_{2}=2.7 \gg \nu_{1}$.
 
%
%
%
%
%
\section{All-order six-photon scattering} \label{sxn:aosps}
As already hinted, six-photon scattering is the dominant process in the generation of higher 
harmonics for $E \ll 1$ in the plane wave set-up we are considering. For this reason we 
choose here to analyse six-photon scattering as the single vacuum interaction. Many of the features 
of the following harmonic spectra will be common to the combined four- and six-photon scattering 
case in \sxnref{sxn:fso}.
\newline

The parameter $v_{2}=\nu_{2}\exp[-(\varphi_{p}/\Phi_{p})^{2}]$ is bounded by $v_{2}\leq\nu_{2}$, so 
the 
different behaviour of the scattered probe will be quantified using the shock parameter $\nu_{2}$. 
As $\nu_{2}$ is increased from zero, two regimes become apparent: i) the perturbative regime 
$\nu_{2} \ll 1$ where the occurrence of higher harmonics is exponentially suppressed; 
ii) the shock regime, where the intensity of the $j$th harmonic is proportional to a power-law 
$j^{\gamma(v)}$, with $\gamma(v)<-2$. 
\newline

 To highlight the nature of the 
harmonic generation 
surrounding shock formation, we refer in the following to the parallel set-up for simplicity, and 
discuss differences in the perpendicular set-up in \sxnref{sxn:poldep}.
\newline

In \figref{fig:triptychA} are log-log plots of three different types of normalised spectrum 
$I(\omega)/I_{p}^{(0)}(\omega_{p})$ in the parallel set-up. In the 
first pane $(a)$ $\nu_{2}=0.05 \ll 1$ and the perturbative regime can be recognised by the exponential 
suppression of higher harmonics. In the middle pane $(b)$ $\nu_{2}=0.6$ and a transition regime can be 
identified in which the lower harmonics are no longer exponentially-suppressed but obey a power-law 
behaviour and the leading-order perturbative expansion is inaccurate for higher harmonics. In the 
final pane $(c)$ $\nu_{2}=1$ and the entire plotted spectrum has a power-law behaviour, distinctive 
of the shock regime, in which an all-order expansion is required to even reach a correct 
qualitative 
conclusion.
Since we are considering only six-photon scattering, we set 
$\nu_{2}=\nu$ and $v_{2}=v$ in the following discussion.
\begin{widetext} 
\begin{center}
\begin{figure}[!h]
\includegraphics[width=17cm]{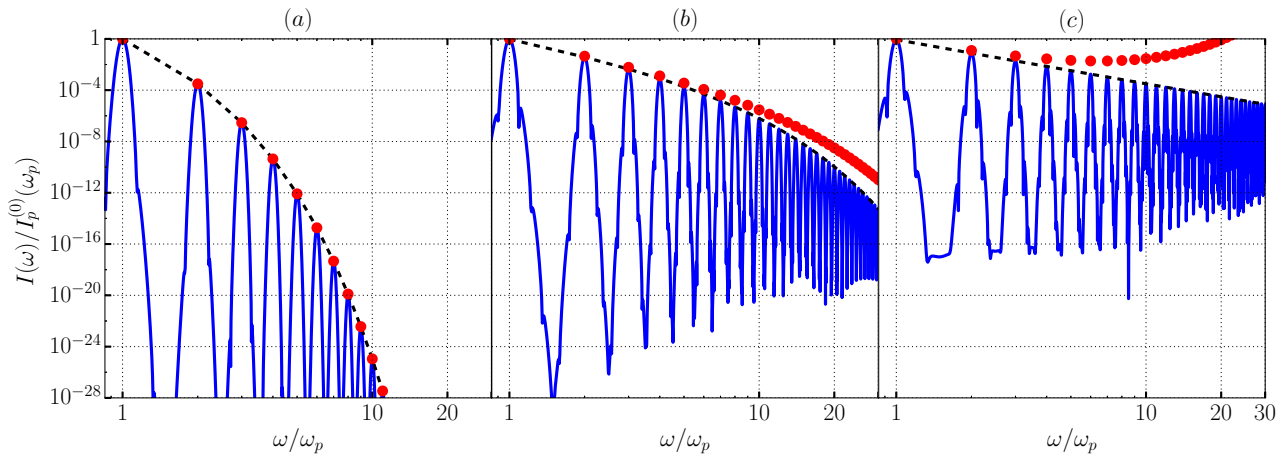}
\caption{(Color online) Harmonic spectra in the parallel set-up for
  different regimes of solution:
$(a)\; \nu_{2}=0.05, (b)\; \nu_{2}=0.6, (c)\; \nu_{2}=1$. The dots show the leading-order perturbative term, the dashed line is 
the 
all-order analytical solution and the solid line is from numerical simulation.} 
\label{fig:triptychA}
\end{figure}
\end{center}
\end{widetext}

\subsection{Perturbative regime}
If $\nu \ll 1$, the amplitude of each harmonic in the scattered electric field is:
\bea
|a_{j}(\nu)| = \frac{1}{\Gamma(1+j)}\left(\frac{\nu j}{2}\right)^{j-1} +\trm{O}(\nu^{j+1}).
\eea
For $\nu j \ll 1$ but $j\gg1$, using Stirling's approximation \cite{arfken12}
$\Gamma(1+j)\approx \sqrt{2\pi j}(j/e)^{j}$, we see:
\bea
|a_{j}(\nu)| = \frac{(\nu\mbox{e})^{j}}{\nu j^{3/2}\sqrt{2\pi}}, \qquad 
\left|\frac{a_{j+1}(\nu)}{a_{j}(\nu)}\right| \approx \nu e,
\eea
and the exponential dependency of each harmonic becomes manifest. In the first pane of 
\figref{fig:triptychA}, the dots denote the intensities of the harmonics when only the leading 
perturbative order is taken into account. The excellent agreement is typical of the perturbative 
regime, in 
which only a small proportion of probe photons have scattered, and double-scattering is much less 
probable than single-scattering. In the transition regime, the 
leading-order terms of the perturbative expansion overestimate the intensity of the higher 
harmonics. In the shock regime, the leading-order perturbation terms both qualitatively and 
quantitatively disagree with the numerical solution and all-order analytical solution.
\newline

\subsection{Shock regime}
In this regime, $\nu$ no longer fulfills $\nu\ll1$ and all orders of the perturbative 
expansion must be 
summed in order to calculate the spectrum of generated harmonics. This is demonstrated in the 
third pane of \figref{fig:triptychA} which shows excellent agreement between the numerical and 
analytical solution \eqnrefs{eqn:Epn}{eqn:harmonicCoeffs}. We note that even though the all-orders 
solution includes the phase-dependent parameter $v=\nu\exp(-(\vphi_{p}/\Phi_{p})^{2})$, we 
can still arrive at a qualitative understanding of this regime by considering the effect on the 
probe pulse at the point $\vphi_{p}=0$. In this case, $v=\nu$ and the relative amplitude of 
consecutive harmonics is
\bea
\left|\frac{a_{j+1}(\nu)}{a_{j}(\nu)}\right| = \left|\frac{J_{j+1}[(j+1)\nu]}{J_{j}(j\nu)}\right| 
\frac{j}{j+1}. \label{eqn:shbess}
\eea
Using the asymptotic form for $j\nu \to \infty$, $|J_{j}(j\nu)| \sim (2\pi j)^{\nicefrac[]{-1}{2}}$ 
(when phase 
terms are neglected) \cite{watson22}, we see that for large enough argument, the 
ratio of harmonic 
amplitudes becomes:
\bea
|a_{j}(\nu)| \sim \frac{1}{\nu\,j^{\nicefrac[]{3}{2}}}\sqrt{\frac{2}{\pi}}, \qquad 
\left|\frac{a_{j+1}(\nu)}{a_{j}(\nu)}\right| \sim \left(\frac{j}{j+1}\right)^{\nicefrac[]{3}{2}},
\eea
and the power-law behaviour is manifest. For $\nu=1$, this gives a
ratio of the intensity of the $j$th harmonic to 
the initial probe intensity, 
$I^{(j)}_{p}(\vphi_{p})/I_{p}^{(0)}(\vphi_{p})=[E^{(j)}_{p}(\vphi_{p})/E^{(0)}_{p}(\vphi_{p})]^{2}$ 
of
\bea
\log \left(\frac{I^{(j)}_{p}(\vphi_{p}=0)}{I_{p}^{(0)}(\vphi_{p}=0)}\right) \underset{\footnotesize 
jv\to\infty}{\sim} 
-\log\frac{2}{\pi} - 3\log j.
\eea
The predicted gradient of $\gamma=-3$ should be an overestimate because for all parts of the probe 
apart from at $\vphi_{p}=0$, $v<\nu$. In fact, the full result in the third pane of 
\figref{fig:triptychA} yields $\gamma=-3.4$.
\newline

\begin{figure}[!h]
\label{shock-par}
\includegraphics[width=0.7\linewidth]{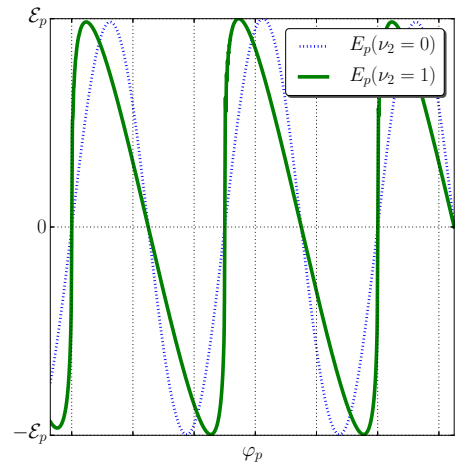}
\caption{(Color online) After passing through the polarised vacuum in the parallel set-up, the probe pulse 
wavefronts can steepen significantly.} \label{fig:shock1}
\end{figure}
A plot of the scattered probe field and induced electromagnetic shock is displayed in 
\figref{fig:shock1}. Those parts of the 
probe field that are 
positive 
and have a larger amplitude are decelerated more than those that are positive with a 
smaller amplitude. Where the field is positive, this leads to a steepening behind the peaks. Those 
parts of the probe that are negative but have a larger amplitude 
are decelerated less than those that are negative but have a smaller amplitude, hence leading to a 
steepening in the opposite direction where the field is negative. The result is the development of 
a saw-tooth waveform shown in \figref{fig:shock1}, which is typical of a second-order 
susceptibility 
\cite{radnor08}.
\newline

\begin{center}
 \begin{figure}[!h]
  \includegraphics[width=6cm]{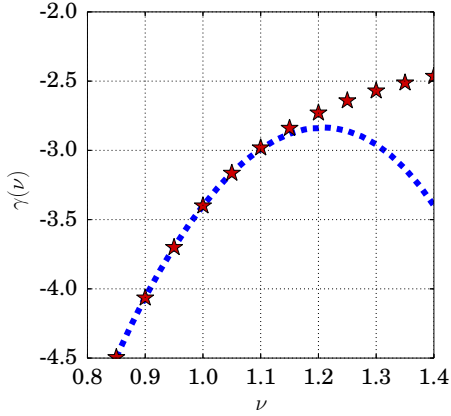}
 \caption{(Color online) Comparison of the power-law exponent $\gamma(\nu)$ for different values of the shock 
parameter $\nu$, as calculated using the fourth and tenth harmonics from the analytical (dashed 
line) and numerical (points) solutions.} \label{fig:gammaPlot}
 \end{figure}
\end{center}

The coefficient of the $j$th harmonic is weighted with the Bessel function
$J_{j}(jv)$. When $v$ is small, $J_{j}(jv)$ is a rapidly decaying function of $j$ so higher 
harmonics are strongly suppressed. As $v\to 1^{-}$, the decay becomes much shallower. So a 
simplified picture of what type of shock is generated for the scenario explored in 
this 
paper can be made by setting the Bessel function to a constant. In the parallel set-up, a 
discontinuous electric 
field with 
a backwards-leaning waveform of the form \figref{fig:shock1} is generated with:
\bea
\mbf{E}(\vphi)=\veps\mathcal{E}\sum_{j=1}^{\infty}(-1)^{j}\left[\frac{\cos(2j-1)\vphi}{
2j-1}+\frac{\sin2j\vphi}{2j}\right], \label{eqn:waveform1}
\eea
with polarisation $\veps$ and amplitude $\mathcal{E}$, and the corresponding intensity spectrum has 
a power law $\sim j^{-2}$ for harmonic $j$. Indeed we find on a plot of $\gamma(\nu)$ (see 
\figref{fig:gammaPlot}), that as $\nu$ increases above $1$, the power-law exponent in the numerical 
spectrum increases, tending toward a theoretical maximum of $-2$, at which point the lack of a 
unique solution to Maxwell's equations would halt further propagation of the probe. For $\nu>1$, 
the numerical spectrum displays a variable power law, which is shallower for higher 
harmonics where the agreement with the analytical solution \eqnrefs{eqn:Epn}{eqn:harmonicCoeffs} 
becomes increasingly worse. The power law exponent calculated using the fourth and tenth harmonic 
is displayed in \figref{fig:gammaPlot}, where unlike in the numerical solution, in which the 
spectrum becomes progressively shallower, the analytical solution reaches a maximum shallowness. It 
is unclear what physical mechanism would cause this maximum to occur, which suggests this is a 
limitation of the viability of the analytical solution. Indeed when 
$v>1$ in the analytical solution, $J_{j}(jv)$ can oscillate with $j$, and the ordering of harmonics 
can become no longer monotonic.
As the numerical spectrum becomes shallower, very high harmonics appear, which questions 
the validity condition $j\omega_{p} \ll m$ for using the Heisenberg-Euler 
Lagrangian to describe the vacuum interaction, and questions how steep the power law can become 
before relaxation processes would take over.

\subsection{Polarisation Dependency} \label{sxn:poldep}
The previous sections are for the parallel set-up. For the perpendicular set-up, even harmonics 
are generated in the parallel mode $\vepspar$ and odd harmonics in the perpendicular 
mode $\vepsperp$. This is demonstrated in the spectrum in \figref{fig:triptychB}, where the thick 
and thin lines distinguish how the generated harmonics are polarised. 
\newline

\begin{figure}[!h]
\includegraphics[width=0.7\linewidth]{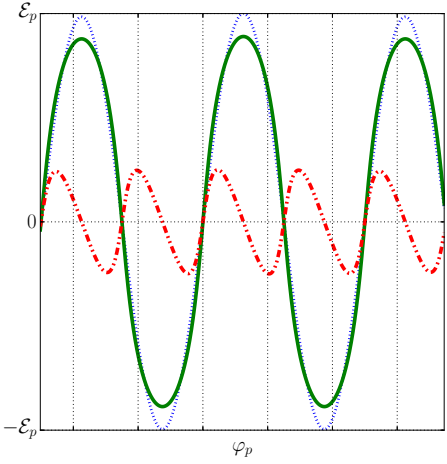}
\caption{(Color online) A probe that is initially polarised perpendicular to the background (blue dashed line) 
experiences different shocks in the $\vepsperp$ (dot-dashed red line) and 
$\vepspar$ (solid green line) modes (color online).} \label{fig:shock2}
\end{figure}

The shock wave generated in the perpendicular set-up is displayed in \figref{fig:shock2}. The 
scattered field in the $\vepspar$ mode demonstrates a shock of a different nature to in the 
parallel set-up, tending towards a square rather than a saw-tooth waveform. Such a waveform can be 
generated with the sum:
\bea
\mbf{E}_{\trm{square}}(\vphi)=\veps\mathcal{E}\sum_{j=1}^{\infty}(-1)^{j}\,\frac{\cos (2j-1)\vphi}{
2j-1}, \label{eqn:waveform2a}
\eea
which is just the odd frequencies of \eqnref{eqn:waveform1}.
\begin{widetext} 
\begin{center}
 \begin{figure}[!h]
 \includegraphics[width=17cm]{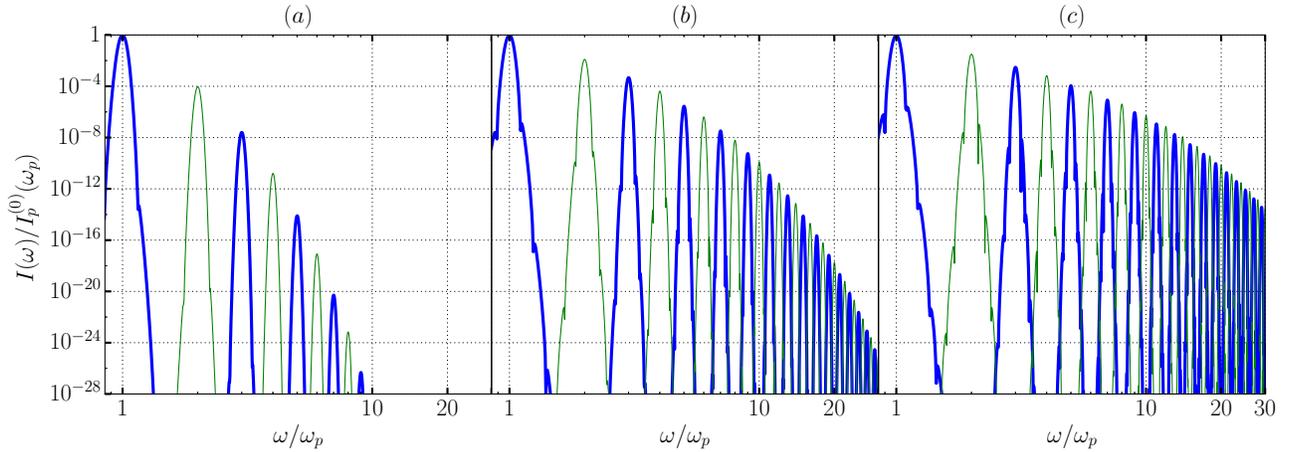}
 \caption{(Color online) Harmonic spectra from numerical simulation of the perpendicular set-up for different 
regimes of solution: $(a)\; \nu_{2}=0.05, (b)\; \nu_{2}=0.6, (c)\; \nu_{2}=1$. The thick blue (thin green) peaks are harmonics 
parallel to the probe (strong) pulse.} \label{fig:triptychB}
 \end{figure}
\end{center}
\end{widetext}
In the $\vepsperp$ mode, a similar shock to in the parallel set-up is seen, 
only with double the frequency. Such a saw-tooth electric 
field is given by the sum \cite{arfken12}:
\bea
\mbf{E}_{\trm{saw}}(\vphi)=\veps\mathcal{E}\sum_{j=1}^{\infty}(-1)^{j}\,\frac{\sin 2j\vphi}{
2j}, \label{eqn:waveform2b}
\eea
which is just the even frequencies of \eqnref{eqn:waveform1}, beginning at double the frequency of 
the seed probe field.
%
%
%
%
\section{All-order four- and six-photon scattering} \label{sxn:fso}
Although six-photon scattering is the most efficient process in generating high harmonics, for 
the parameter regime we are interested in, the effect of four-photon scattering as a modified 
vacuum refractive index cannot be neglected. Since the interaction with the vacuum includes powers 
of the probe field, the effects of phase lag and harmonic 
generation can mix 
in a highly nonlinear way. In this section we give the results of numerical simulations that 
include 
both processes.
\newline

For the parallel set-up, the spectrum 
generated by six-photon scattering (for example, as shown in \figref{fig:triptychA}), is 
not visibly affected by the inclusion of four-photon scattering. However, for the perpendicular 
set-up, since 
even and odd harmonics are in different polarisation modes and since the vacuum is birefringent so 
each polarisation mode experiences a different phase lag, the inclusion 
of 
four-photon scattering was found to increase the asymmetry between the even and odd harmonics 
compared with the purely six-photon scattering case. This is demonstrated in 
\figref{fig:spec_both} for the case $\upsilon_{1}=100$, $\nu_{2}=1$, which compares the spectrum of 
harmonics generated when: i) only 
four-photon scattering is included (left-hand pane); ii) only six-photon scattering is 
included (middle pane) and iii) four- and six- photon scattering are included (right-hand pane). 
The right-hand pane demonstrates the increased asymmetry between even and odd harmonics.

\begin{widetext}
\begin{center}
\begin{figure}[!h]
\includegraphics[width=17cm]{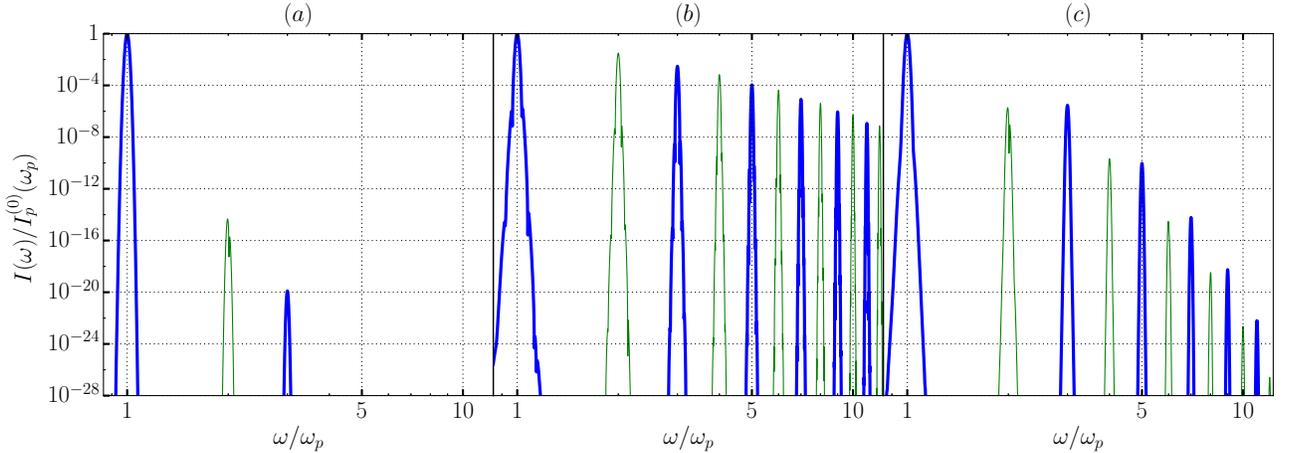}
\caption{(Color online) High harmonic generation for the perpendicular set-up when four- and six-photon scattering 
are present and four-photon scattering is much more prevalent than six-photon scattering 
($\upsilon_{1} = 100$, $\nu_{2}=1$). The first pane $(a)$ is for just
four-photon scattering, the second pane $(b)$
for just six-photon scattering and the third pane $(c)$ for when both are
present. The thick blue (thin green) peaks are again harmonics 
parallel to the probe (strong) pulse.
} \label{fig:spec_both}
\end{figure} 
\end{center}
\end{widetext}

As the case of four- and six-photon scattering differs from the six-photon scattering case 
only for the perpendicular set-up, we focus our discussion on this. Then there are three cases of 
interest: i) weak dispersive: $\upsilon_{1} \ll \nu_{2}$; ii) dispersive: $\upsilon_{1} \approx 
\nu_{2}$; iii) strong dispersive: $\upsilon_{1} \gg \nu_{2}$. The first case of weak vacuum 
dispersion is within the parameter regime of interest, but outside of the regime that can be 
numerically simulated as it would require $\mu_{2}\Phi \gtrsim 0.1$ if the hierarchy 
$\amps\gg\ampp$ 
were to be maintained. In the limit of vanishing dispersion, we expect the results from purely 
six-photon scattering case to be valid (this will be seen to be implied from the results of a 
dispersive vacuum).

\subsection{Dispersive vacuum $\upsilon_{1}\approx\nu_{2}$}
\begin{figure}[!h]
\includegraphics[width=0.7\linewidth, 
draft=false]{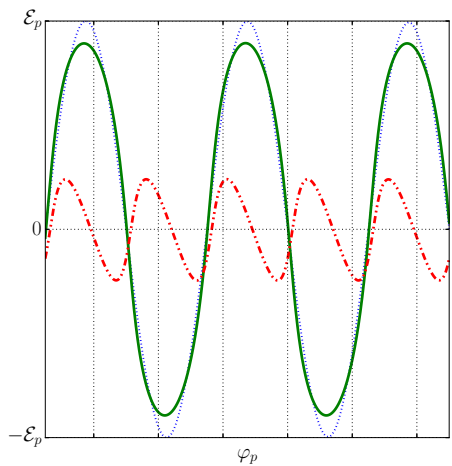}
\caption{(Color online) The weakly-dispersive case for the perpendicular set-up. The initially $\vepspar$ 
polarised probe (blue dashed line) experiences the mixture of the probe-independent vacuum 
refractive index (here $\upsilon_{1}=1$) and the shock-inducing probe-dependent vacuum refractive 
index (here $\nu_{2}=1$). The $\vepspar$ mode (dot-dashed red line) and $\vepsperp$ mode (solid 
green line) behave differently.} 
\label{fig:shock_fso_A}
\end{figure}

\begin{figure}[!h]
\includegraphics[width=0.7\linewidth]{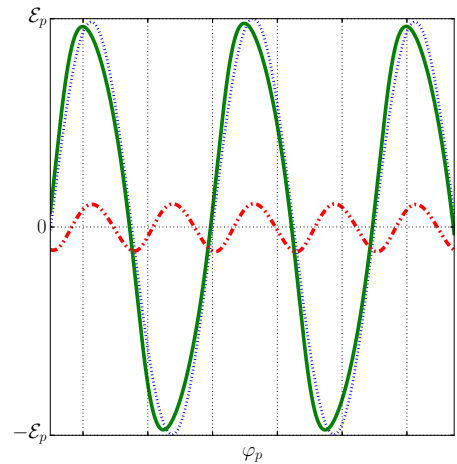}
\caption{(Color online) When the dispersion is increased ($\upsilon_{1}=5$, $\nu_{2}=1$), a different type of 
shock is formed in the $\vepspar$ mode (dot-dashed red line) and the shock in the $\vepsperp$ mode 
(solid green line) is reduced, where the initially $\vepspar$ polarised probe is plotted by the 
blue dashed line.} 
\label{fig:shock_fso_B}
\end{figure}

When vacuum dispersion is significant, one might expect the nature of the shock wave to 
change. Two cases were simulated: i) when $\upsilon_{1}=\nu_{2}=1$ and ii) when $\upsilon_{1}=5$, 
$\nu_{2}=1$. For the first case of equal parameters, the shock wave in \figref{fig:shock_fso_A} was 
generated. This bears a close resemblance to the shock wave generated in the perpendicular set-up 
for a dispersionless vacuum ($\upsilon_{1} \to 0$), i.e. when only six-photon scattering is 
present, but with a noticeable lag due to the now non-unitary refractive index. 
However, when the amount of dispersion is increased, setting $\upsilon_{1}=5$ and 
$\nu_{2}=1$, the shock wave takes on the different form shown in \figref{fig:shock_fso_B}.
In this dispersive case, the parallel mode develops a shock reminiscent of an optical Kerr medium, 
in which the 
polarisation contains a cubic nonlinearity $P_{i} = \chi^{(1)}_{ij}E_{j} + 
\chi^{(3)}_{ijkl}E_{j}E_{k}E_{l}$. 
This is in some ways unsurprising because the parallel mode only contains odd harmonics and 
therefore odd powers of the field, and the 
largest nonlinear term originates from an $E_{p}^{3}$ term. Therefore the symmetry of the 
scattered field when the field direction is swapped $\ampp\to-\ampp$ is different for the parallel 
field (which contains only even powers of $\ampp$) and the perpendicular field (which contains only 
odd powers of $\ampp$).
\newline

Carrier-wave shocking also occurs in nonlinear optical materials. Our findings are similar 
to those reported in \cite{kinsler07}, where excellent agreement was obtained between theory and 
simulation in the 
dispersionless limit of a Kerr-like nonlinear material, but where it was noted how involved
the analysis becomes if there is a complicated phase dependency between the generated harmonics. 
In the current work, in 
the parallel set-up with dispersion (i.e. four- and six-photon scattering present), all harmonics 
experience the same refractive index so a shock wave can build up. In the perpendicular set-up, the 
refractive index in the $\vepsperp$ mode is different to in the $\vepspar$ mode. We are 
studying a regime in which harmonics are generated by a chain of scattering processes. 
Since, in each chain of processes that lead to the generation of a specific harmonic, the probe 
spends a different amount of time in the 
$\vepsperp$ than in the $\vepspar$ mode, the probability for each chain will be multiplied by a 
different phase. When the probability of all possible chains is summed over, it is reduced 
compared to the parallel set-up due to each probability being added incoherently. This leads to a 
suppression of shock wave generation.

\subsection{Strongly-dispersive vacuum $\upsilon_{1}\gg\nu_{2}$}
\begin{figure}[!h]
\includegraphics[width=0.7\linewidth]{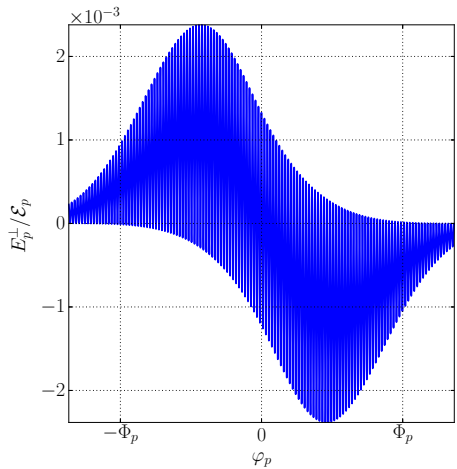}
\caption{(Color online) The probe pulse after having scattered in the strong background when
$\upsilon_{1}=100$, $\nu_{2}=1$.} 
\label{fig:shock_100_1}
\end{figure}

To investigate shock wave generation in the strongly-dispersive regime, we set 
$\upsilon_{1}=100$ and $\nu_{2}=1$. A new type of behaviour becomes apparent, namely the 
deformation of the probe pulse envelope. The bandwidth of the probe is of the order $1/\tau_{p}$ 
but due to dispersive effects, frequencies of this magnitude can no longer be neglected. Since 
$\upsilon_{1}=\delta \vphi_{p} = \omega_{p} T$, where $T$ is the duration of propagation, 
frequencies from the probe envelope separated by $1/\tau_{p}$ will acquire a temporal separation 
relative to the duration of the pulse of $\upsilon_{1}/\omega_{p}\tau_{p} \not\ll 1$. Furthermore, 
the second harmonic is considerably suppressed when dispersion is included, such 
that it is of the same order of magnitude as the scattering of the probe envelope frequency. For 
this reason, the effect on the probe envelope can be seen so clearly in the 
$\vepsperp$ component in \figref{fig:shock_100_1}.

%
%
%
%
\section{Discussion}
\subsection{Comparison with high harmonic generation in oscillating plasmas}
There is a certain similarity between high harmonic generation due to the 
relativistic movement of electrons in the plasma of laser-irradiated foil 
experiments and the virtual electron-positron ``plasma'' of the 
laser-irradiated vacuum. The vacuum is transparent when the invariants $E^{2}-B^{2}$ and 
$\mbf{E}\cdot\mbf{B}$ are zero. Therefore the vacuum is transparent to a pure plane wave 
and these invariants also typically remain much smaller for a single focused pulse than for 
counterpropagating pulses. So unlike with the plasma present in 
a 
foil, the vacuum plasma is first ``activated'' by being polarised by some second ``pump'' 
pulse, similar to in a pump-probe experiment. In the current work, the vacuum was 
polarised by a 
background with the profile of a rectangular function. As the leading-order nonlinear polarisation 
was 
proportional to the applied field cubed, it 
suggests 
that the local charge density is also non-zero in this region. The rectangular function is 
used to 
model 
the electron density in a solid before it is exposed to a laser pulse \cite{mckenna13} and also to 
represent the laser's profile and in capillary discharge waveguides \cite{leemans14}. The 
difference with the vacuum is that the polarised material can in some way be ``formed'' by the pump 
pulse in the moment it is traversed by a probe. 
\newline

For the parallel set-up, all harmonics were generated in the parallel mode, but for the 
perpendicular set-up odd harmonics were generated in the parallel mode with even harmonics 
in the perpendicular one. Just as in single nonlinear Compton scattering \cite{king14c}, the 
generation of the parallel mode is more probable than the perpendicular one. The relationship 
between polarisation and harmonic order is reminiscent of selection rules for harmonics generated 
in 
laser-foil experiments, for example in the ``p-polarised'' (parallel to plane of 
incidence) and 
``s-polarised'' (perpendicular to plane of incidence) harmonics in the widely-used oscillating 
mirror model \cite{lichters96}.
\newline

In the harmonic spectrum generated by a real plasma in laser-foil experiments, there 
is also a region of power-law 
decay and a region of exponential decay, as found here for vacuum high harmonic generation.  For the 
oscillating mirror model, power-law exponents of $\gamma=\nicefrac[]{-5}{2}$ 
\cite{gordienko04} and 
$\gamma=\nicefrac[]{-8}{3}$ \cite{baeva06} have been postulated, and experiments on solid 
targets have recorded intensity-dependent power-law exponents, for example in \cite{norreys96} of
$-5.50<\gamma<-3.38$. These values are close to the analytical and numerical 
values found in the 
current work for vacuum high harmonic generation in the shock regime, $-4.5\lesssim\gamma<-2$ (the 
lower limit corresponds to the gradient when the power-law behaviour becomes manifest at 
$\nu_{2}\approx0.85$). Moreover, 
the power-law exponent $\gamma=\gamma(\nu_{2})$ is also a function of the 
shock parameter $\nu_{2}\propto\Phi$ and therefore increases with further propagation of the probe 
through the polarised vacuum, up to a theoretical maximum of $\gamma(\nu_{2})<-2$. In contrast to 
the overdense plasma 
case, with our plane-wave model and increasing shock parameter, we found no indication of a 
frequency cutoff, although at some frequency, pair-creation processes will play a role. By this we 
mean that higher harmonics can seed tunnelling pair creation in 
the background field \cite{reiss62, nikishov64} or colliding 
photons with wavevectors $k_{1}$ and $k_{2}$ satisfying $k_{1}k_{2}\gtrsim 2m^{2}$ would 
lead to multi-photon pair creation (the Breit-Wheeler process \cite{breit34, pike14}). This would 
presumably deplete the higher harmonics that are directly related to the steepening of the wave 
fronts and act as a wave-breaking mechanism for the shock wave.
\newline

In the parallel set-up, each harmonic has a regular phase relationship to the others and 
so a shock wave can build up as the amplitude of higher harmonics increases. In contrast to this, 
in the perpendicular set-up, since there are many different chains of processes that can lead to 
the creation of a given harmonic, and since in each chain a different amount of time is spent in 
each polarisation, which leads to different dispersion relations, the phase of each 
harmonic is related to the others in a non-trivial way and they are summed incoherently. This 
behaviour is similar to that found in 
studies of non-linear optical materials \cite{kinsler07}, and leads to the suppression of shock 
wave generation.
\newline

Although high harmonic generation is present in laser-gas and laser-liquid experiments, the 
spectrum generated is of a completely different form. As harmonics are generated via 
the three-step recombination mechanism undergone by an electron in the Coulomb field of a 
nucleus, the electrons' trajectory and hence harmonics generated, are of a fundamentally different 
nature and demonstrate a genuine ``plateau'' region in spectra that is not present in 
vacuum high harmonic generation as studied in the present work \cite{winterfeldt08}.

\subsection{Validity of approach}
By considering colliding plane waves, scattering in the transverse direction was ignored. One can 
estimate when this is a good approximation by defining the diffraction parameter $l = 
\trm{w}^{2}/\lambda_{p}\tau_{s}$, where $\trm{w}$ is the width of the probe pulse in the transverse 
plane (assumed smaller than the width of the background). 
When $l\gg 1$ one is in the ``near zone'' and diffraction effects should be negligible whereas 
$l\ll 
1$ represents the ``far zone'' and diffraction effects become important \cite{levi68}.
\newline

The numerical simulation and analytical calculation predict a self-steeping of the probe 
wavefronts, which increases with shock parameter $\nu$, until the 
wavefronts reach a theoretical maximum of becoming infinitely 
steep at which point the solutions to the wave equation are no longer unique. Since the 
Heisenberg-Euler Lagrangian is expected to be valid when the typical scale of a field inhomogeneity 
is much larger than the reduced Compton wavelength, this infinite steepening is not expected to be 
physically realisable. Moreover, no relaxation processes are included. If transverse dimensions 
would be taken into account, since six-photon scattering depends on the probe amplitude, 
self-focusing effects should be present. 
Furthermore, self-focusing can also occur via four-photon scattering as the probability for 
asymptotic 
second-harmonic generation via four-photon scattering becomes non-zero when the colliding probe 
photons do not propagate in parallel. So when transverse dimensions are included, as 
the probe propagates, it becomes less like a plane wave and the higher harmonics can seed real 
electron-positron pair creation as previously described.
\newline

The polarisation of other vacuum virtual particle species such as muons, pions and quarks 
was neglected, as the energy scale associated with these particles is much 
higher \cite{ferreira08}. For that reason, we confined our discussion to the polarisation of 
virtual electron-positron pairs.

\subsection{Measurability}
Vacuum high harmonic generation in the shock regime becomes important when the shock parameter $\nu 
\approx 1$. Taking as an example six-photon scattering for the parallel set-up,
$\nu=\nu_{2}=192\mu_{2}\mathcal{E}_{s}^{3}\mathcal{E}_{p}\Phi$. The current record for 
the highest electric field of a laser pulse produced in a laboratory \cite{yanovsky08} is of the 
order $3\times 10^{-4} \Ecr$. Recalling that fields are written in units of the critical field, and 
that $\mu_{2} = \nicefrac[]{\alpha}{315\pi} \ll 1$, it is clear that the shock regime is currently 
well out of the 
reach of optical laser-based experiments. Vacuum polarisation effects that can more 
likely be measured in 
laser-based experiments include elastic photon-photon scattering 
\cite{mendonca06,heinzl_birefringence06,tommasini07,tommasini10,king10a,king10b,hatsagortsyan11,
monden11,dinu14a,dinu14b,hu14b,karbstein14a,karbstein15} or lowest-order photon merging 
\cite{marklund_PRL_01,lundstroem_PRL_06,king12,gies14}. (A review of strong-field QED effects can be 
found in \cite*{marklund_review06, dipiazza12}.) The current best experimental limits for 
photon-photon scattering in an all-optical laser set-up \cite{bernard00} and combining magnetic 
fields with resonant optical cavities \cite{pvlas12, rizzo13} are still orders of magnitude above 
the QED prediction.
\newline

Where such vacuum electromagnetic shocks and accompanying harmonic generation might play a 
role, is in the evolution of X-ray pulsars and strongly-magnetised neutron stars or ``magnetars'' 
\cite{mazets79,duncan92,ho03}. Photons are emitted from the surface of such objects and propagate 
through magnetic fields of strength up to and beyond $\Ecr$ (in the system of
units we use, $B_{\trm{cr}}=\Ecr$), in plasmas of around 
$0.1$-$10\,\trm{cm}$ in depth \cite{harding06}. The current results were derived for a constant 
crossed field background, but can be generalised to a constant magnetic field, which should be a 
good approximation to the local field in strongly-magnetic pulsars, which is expected to be 
that of a dipole \cite{harding06} on the stellar scale.

\section{Conclusion}
When the quantum nature of the vacuum is taken into account, an electromagnetic shock 
accompanies high harmonic generation in an oscillating plane probe pulse 
counterpropagating through a stronger slowly-varying plane pulse. We have identified a 
nonlinear shock 
parameter that indicates when the self-interaction of the probe due to the polarised vacuum becomes 
important and have shown that this can be consistently described using a probe-dependent vacuum 
refractive index.

As the shock parameter increases from zero, the spectrum of generated harmonics moves from an 
exponential decay to a power-law decay. The intensity of the $j$th harmonic in the shock regime was 
found in an all-order analytical solution and numerical simulation to be $j^{\gamma}$, where 
$\gamma$ 
increases with propagation distance. A power law behaviour was observed for 
$-4.5\lesssim\gamma\lesssim -2.4$, where the exponent is theoretically limited by $\gamma<-2$ as 
the probe pulse wavefronts would become infinitely steep and could no longer propagate. 
Due to the very high 
generated frequencies, the Heisenberg-Euler approach is no longer applicable at this point. 
Moreover, relaxation processes such as photon-seeded and Breit-Wheeler pair creation should 
then become probable.

When the polarisation of the probe and background is parallel, all harmonics are generated in the 
parallel mode, but when the probe is perpendicularly-polarised to the background, odd and even 
harmonics are split into probe and background polarisation modes 
respectively. Due to the birefringence of 
the vacuum, the probe polarisation mode is generated more abundantly 
than in the background polarisation mode. 
Moreover, due to the separation of frequencies, the parallel set-up displays a saw-tooth shock in 
the parallel mode, whereas the perpendicular set-up displays a Kerr-like shock.

Both the simulational and analytical methods presented can be generalised to more complicated probe 
and background fields.
%
%
%
%
\section{Acknowledgements}
P. B. acknowledges the very useful advice of A. Hindmarsh during development of the 
computational 
simulation. B. K. acknowledges the hospitality of H. R. and the Arnold Sommerfeld Center 
for Theoretical Physics at the Ludwig Maximilian University of Munich.  This work was partially 
funded by Deutsche Forschungsgemeinschaft DFG
under contracts FOR968, RU633/1-1, SFB-TR18 project B12 and EXC-158
(cluster of excellence MAP). Plots were generated with {\tt{Matplotlib}} 
\cite{matplotlib}.
%
%
%
%
\bibliography{current}
%
%
%
%
\appendix
\section{Coefficients for modified Maxwell Equations} \label{sxn:Cs}
We define $\mathcal{L}_{xy}=\partial^{2}\mathcal{L}_{\trm{HE}}/\partial
x\partial y$ and $r^{2}=a^{2}+b^{2}$.
\begin{align}
  C_{1} &= 4\pi\frac{a\mathcal{L}_{a}- b\mathcal{L}_{b}}{r^{2}},\\
  C_{2} &= \pi\frac{1}{r^{6}}\,  \left[a(a^{2}-3b^{2})\mathcal{L}_{a}
+b(b^{2}-3a^{2})
\mathcal{L}_{b}
  \right.\nonumber\\
  &\qquad\left.-r^{2}\left(a^{2}\,\mathcal{L}_{aa} -
2ab\,\mathcal{L}_{ab} +
b^{2}\mathcal{L}_{bb}\right)\right]\\
  C_{3} &= \pi \frac{1}{r^{6}}\,
\left[a(3b^{2}-a^{2})\mathcal{L}_{a}+b(3a^{2}-b^{2})
    \mathcal{L}_{b}\right.
  \nonumber\\
  &\qquad\left.-r^{2}\left(b^{2}\,\mathcal{L}_{aa} +
2ab\,\mathcal{L}_{ab} +
a^{2}\,\mathcal{L}_{bb}\right)
  \right]\\
  C_{4} &= \pi\frac{1}{r^{6}}\,
  \left[b(3a^{2}-b^{2})\mathcal{L}_{a}+a(a^{2}-3b^{2})
    \mathcal{L}_{b}\right.
  \nonumber\\
&\qquad\left.-r^{2}\left(ab\,\mathcal{L}_{aa}
      +(a^{2}-b^{2})\,\mathcal{L}_{ab} - ab\,\mathcal{L}_{bb}\right)
  \right]\label{eqn:coeffs_maxwell}
  \end{align}
For the first order (box diagram) and the second order (hexagon diagram)
in the weak-field
expansion, we find the following coefficients:
\begin{align}
C_{1,\text{Box}}&=\frac{2\alpha}{45 \pi}(E^2-B^2) & \quad \label{eqn:C1b}
C_{2,\text{Box}}&=-\frac{\alpha}{45\pi}\\
C_{3,\text{Box}}&=\frac{7}{4} C_{2,\text{Box}}  &\quad C_{4,\text{Box}}&= 0
\end{align}
\begin{align}
C_{1,\text{Hex}}&=\frac{2\alpha}{315 \pi}[6(E^2-B^2)^2+13 (\mbf{E}\cdot
\mbf{B})^2]\\
C_{2,\text{Hex}}&=-\frac{4\alpha}{105\pi}(E^2-B^2)\\
C_{3,\text{Hex}}&=\frac{13}{24} C_{2,\text{Hex}}\\
C_{4,\text{Hex}}&= -\frac{13\alpha}{315\pi} | (\mbf{E}\cdot\mbf{B})| \label{eqn:C4h}
\end{align}
%
%
%
%
\section{Matrix form of modified Maxwell Equations}
\label{sec:matricesAB}
The modified Maxwell equations \eqnrefs{eqn:MW1}{eqn:MW2} can be written in matrix form:
\begin{align}
\left(\mathbbm{1}_{4}+\mbf{X}\right)\partial_{t}\mbf{f}+\left(\mathbf{Q}+\mbf{Y}\right)\partial_{z}
\mbf{f}=0,
\end{align}
where $\mbf{f} = (E_{x},E_{y},B_{x},B_{y})^{T}$, $\mathbbm{1}_{4}$ is the identity in four 
dimensions, $\mbf{Q}=\trm{adiag}(1,-1,-1,1)$ and $\mbf{X}=(x_{ij})$, 
$\mbf{Y}=(y_{ij})$ are the vacuum perturbation, where the non-zero elements are given by: 
\begin{flalign*}
x_{11}&= C_1 - C_2 \rho_{11}- C_3 \rho_{33} - 2 C_4 \rho_{13} \\
x_{12}&= - C_2 \rho_{12}-C_3 \rho_{34} - C_4 (\rho_{14} +  \rho_{23}) \\
x_{13}&= (C_2 - C_3) \rho_{13}+ C_4 (\rho_{33} - \rho_{11}) \\
x_{14}&=  C_2 \rho_{14} - C_3 \rho_{23}  + C_4 (\rho_{34} -\rho_{12}) \\
x_{21}&= - C_2 \rho_{12}- C_3 \rho_{34}- C_4 (\rho_{14} + \rho_{23}) \\
x_{22}&= C_1 - C_2 \rho_{22}- C_3 \rho_{44}- 2 C_4 \rho_{24}\\
x_{23}&= C_2 \rho_{23} - C_3 \rho_{14} + C_4 (\rho_{34} - \rho_{12})\\
x_{24}&=  (C_2  - C_3) \rho_{24}+C_4 (\rho_{44} - \rho_{22}) \\
y_{11}&= -C_2\rho_{14}+ C_3\rho_{23} + C_4 (\rho_{12}- \rho_{34})\\
y_{12}&= -(C_2 - C_3) \rho_{24} +C_4 (\rho_{22}-\rho_{44})\\
y_{13}&= C_2  \rho_{34} + C_3 \rho_{12} - C_4 (\rho_{14}+\rho_{23})\\
y_{14}&= C_1 + C_2 \rho_{44}+ C_3\rho_{22}- 2 C_4 \rho_{24}\\
y_{21}&=(C_2 -C_3)\rho_{13}+C_4 (\rho_{33}- \rho_{11})\\
y_{22}&= C_2 \rho_{23} - C_3 \rho_{14}+C_4 (\rho_{34} - \rho_{12})\\
y_{23}&= -C_1 - C_2 \rho_{33} - C_3 \rho_{11}+ 2 C_4 \rho_{13}\\
y_{24}&= - C_2 \rho_{34} - C_3 \rho_{12}+ C_4 (\rho_{14} +\rho_{23} )
\end{flalign*}
where we define $\rho_{ij}:=4 f_i f_j$, such that e.g. $\rho_{14} = 4 E_x B_y$.


\section{Biased finite differences}
\label{sec:stencils}
The action of the matrix $\mathbf{D}$ on the vector $\mathbf{\tilde u}$ can
be encoded in the use of an adaption of the DSS020 function from
\cite{schiesser1991numerical}:
\begin{align*}
\mathbf{D}\mathbf{\tilde u}=
\begin{pmatrix}
d_-(u_1^1)\\
d_-(u_2^1)\\
d_+(u_3^1)\\
d_+(u_4^1)\\
d_-(u_1^2)\\
d_-(u_2^2)\\
d_+(u_3^2)\\
d_+(u_4^2)\\
\vdots
\end{pmatrix}
\end{align*}
where the function $d_-(u^l)$ is defined as 
\begin{align*}
d_-(u^l):=\\
  l=1:&\\
  &q(-25 u^1 + 48 u^2 - 36 u^3 + 16 u^4 - 3 u^5)\\
  l=N-2:&\\
  &q(u^{N-4} - 8 u^{N-3} + 8 u^{N-1} - u^N)\\
  l=N-1:&\\
  &q(- u^{N-4} + 6 u^{N-3} - 18 u^{N-2}+ 10  u^{N-1} + 3 u^{N})\\
  l=N:&\\
  &q(3 u^{N-4} - 16 u^{N-3} + 36 u^{N-2} - 48 u^{N-1} + 25 u^{N})\\
  \trm{else}:&\\
   &q(-3 u^{l-1} -10 u^{l} + 18 u^{l+1} -6 u^{l+2}
   + u^{l+3})
 \end{align*}\\
 with $q= \nicefrac{1}{12\Delta z}$ and $d_+(u^l)$ as
\begin{align*}
d_+(u^l):=\\
  l=1:&\\
  &q(-25 u^1 + 48 u^{2} -36 u^{3} + 16 u^{4} - 3 u^{5})\\
  l=2:&\\
  &q(-3 u^{1} -10 u^2 + 18 u^{3} -6 u^{4} + u^{5})\\
  l=3:&\\
  &q( u^{1} - 8 u^{2} + 8 u^{4} - u^{5})\\
  l=N:&\\
  &q( 3u^{N-4} - 16 u^{N-3} +36 u^{N-2}  -48 u^{N-1} + 25 u^{N})\\
  \trm{else}:&\\
   &q(-u^{l-3} + 6 u^{l-2} -18 u^{l-1} +10 u^l + 3 u^{l+1} ).
\end{align*}


\end{document}